# Interactions between a propagating detonation wave and circular water cloud in hydrogen/air mixture

Yong Xu, Huangwei Zhang[*]

*Department of Mechanical Engineering, National University of Singapore, 9 Engineering Drive 1,*

*Singapore, 117576, Republic of Singapore*

**Abstract**

Interactions between a propagating hydrogen/air detonation wave and circular water cloud are studied. Eulerian-Lagrangian method involving two-way gas-droplet coupling is applied. Different droplet (diameter, concentration) and cloud (diameter) properties are considered. Results show that droplet size, concentration and cloud radius have significant effects on peak pressure trajectory of the detonation wave. Three propagation modes are identified: perturbed propagation, leeward re-detonation, and detonation extinction. Leeward re-detonation is analyzed from unsteady evolutions of gas and liquid droplet quantities. The detonation is re-initiated by a local hot spot from shock focusing of upper and lower diffracted detonations. Disintegration of water droplets proceeds when the detonation wave crosses the cloud. In addition, detonation extinction is featured by quickly fading peak pressure trajectories when the detonation wave passes the larger cloud, and no local autoignition occurs in the shock focusing area. Evolutions of thermochemical structures from the shocked area in an extinction process are also studied. The transfer rates of mass, energy and momentum of detonation success and failure are analyzed. Moreover, parametric studies demonstrate that the critical cloud size to quench a detonation decreases when the droplet concentration is increased. However, when the droplet concentration is beyond 0.84 kg/m$^3$, the critical cloud size is negligibly influenced due to small droplets. Two-phase fluid interfacial instability is observed, and the mechanism of cloud evolution is studied with the distributions of droplet, vorticity, density / pressure gradient magnitudes, and gas velocity. Analysis confirms that velocity difference (due to two-phase momentum exchange) dominates the formation of large-scale vortices from the southern and northern poles, corresponding to Kelvin–Helmholtz instability. Moreover, the influences of effective Atwood number on the evolutions of cloud morphology, vorticity, and gas velocity are evaluated. Results show that higher droplet concentration results in wider droplet dispersion range due to larger vortices.

---

[*] Corresponding author. Tel.: +65 6516 2557; Fax: +65 6779 1459.
*E-mail address*: huangwei.zhang@nus.edu.sg.

# 1. Introduction

Hydrogen ($H_2$), as a clean fuel, has been deemed a promising solution to reduce carbon emissions from fossil fuel combustion. Nonetheless, $H_2$ leakage, e.g., from pressurized tanks, may cause ignition and even detonation, due to its low ignition energy and wide flammability limit. Therefore, inhibition of $H_2$ explosion hazards is crucial to materialize its wide applications in the foreseeable future. Water spray with fine droplets is an ideal inhibitor for gas explosions [1], because it can absorb heat from gas phase due to large heat capacity, latent heat of evaporation and specific surface area.

Numerous studies about shock or blast attenuation by water sprays have been reported, as summarized in Ref. [2]. For instance, Jourdan et al. [3] use water aerosol shock tube experiments to study shock attenuation in a cloud of water droplets, and correlate the degree of shock attenuation with shock tube and droplet properties. Using the similar facilities, Chauvin et al. [4] find the peculiar pressure evolution after the transmitted shock in two-phase mixture. They also observe that the overpressure peak inside the water cloud decreases with increased cloud height and decreased shock intensity. Adiga et al. [5] unveil the physical picture of fine water droplet breakup upon shock passage, and they find that the evaporation energy is important in weakening the shock, whilst the droplet breakup energy is secondary. Moreover, Eulerian−Lagrangian simulations are performed by Ananth et al. [6] to examine the effects of fine water mists on a confined blast. It is concluded that the latent heat absorption is dominant for blast mitigation, followed by convective heat transfer and momentum exchange. Schwer and Kailasanath [7] simulate unconfined explosions in water sprays, but find that momentum extraction by water droplets plays a more important role than droplet evaporation in weakening the blast. Shibue et al. [8] numerically study the blast-mitigation effect resulting from the quasi-steady drag force between the shocked air and stationary water droplets. Their results demonstrate that higher momentum loss of the air results in greater blast inhibition.

There are also studies on the performance of water sprays in terms of detonation inhibition. Thomas et al. [9], for example, attribute detonation failure to high heat loss by water droplets, and water droplet diameter and loading are identified as the key factors. Niedzielska et al. [10] also observe that small droplets have strong influence on quenching a detonation due to their fast evaporation rate. Jarsalé et al. [11] find that water droplets do not alter the ratio of hydrodynamic thickness to detonation cell size, but can influence the detonation stability. Moreover, a global reduction of the average cell size with equivalence ratio is observed, based on a given water mass fraction. Besides the foregoing experimental work, Watanabe et al. [12] observe from their simulations that the cellular structures of hydrogen detonations with water sprays become more regular, compared to the droplet-free detonations. Their results also show that droplet breakup mainly occurs near the shock front, and the average diameter of the disintegrated water droplets is independent on the initial propagation speed of the shock front [13]. Shi et al. [14] simulate the effects of a water curtain on incident methane detonation and reveal that the water curtain not only weakens the incident detonation wave, but also prevent re-ignition of the quenched detonation. Moreover, their results show that the convective heat transfer by water droplets plays a significant role in quenching a detonation.

More recently, Xu and Zhang [15] study the unsteady phenomena in hydrogen/oxygen/argon detonation with water mists based on a one-dimensional configuration. They find that water mists result in detonation galloping motion and even failure, but these effects depend on droplet properties. Furthermore, propagation regimes (i.e., failure or successful transmission) of two-dimensional hydrogen/air detonations in water mists are predicted in parameter space of droplet loading and diameter by Xu et al. [2]. They also analyze the evolutions of the characteristic chemical structure between the reaction front and shock front for both propagating and failed detonations. In addition, it is found that the interphase energy and momentum transfer have more direct influence on the induction zone than water droplet evaporation.

In most preceding studies (e.g., [2]-[15]), sprayed water is considered as a flooding species in the entire studied domain. However, such idealized distributions would be practically rare, considering relatively long droplet dispersion timescale compared to that of detonation wave (DW). Instead, water droplets for explosion suppression are largely generated from sprinklers, which lead to small-range water spray cloud when they are activated. However, how a DW interacts with a localized water cloud and the configuration of water cloud (such as the cloud size or spray droplet diameter) for detonation inhibition have not been fully understood yet.

In this work, hydrogen/air detonations interacting with a circular water spray cloud are simulated. This problem essentially embodies a wealth of interesting physico-chemical processes, e.g., detonation/shock diffraction and refraction, detonation extinction and re-initiation, droplet breakup and vaporization, shock focusing, as well as multiphase fluid interfacial instability. The Eulerian-Lagrangian method considering two-way gas − liquid coupling will be used. The overarching objectives of our studies are to clarify: (1) the effects of water cloud / droplet properties on incident detonation dynamics; and (2) how the shocked water cloud evolves with fluid interfacial instability behaviours. The results from this work will be useful to provide scientific evidence for mechanism and performance of hydrogen detonation inhibition with water sprays, as well as for fundamentals of interfacial instability in two-phase flows. The rest of the manuscript is structured as below. The mathematical model and numerical method will be presented in Section 2, whilst the physical model and numerical implementation will be clarified in Section 3. The results and discussion will be detailed in Section 4, followed by main conclusions in Section 5.

## 2. Mathematical model

The Eulerian-Lagrangian method is used to simulate the gas—droplet two-phase compressible reactive flows. The details of the mathematical model can be found in our recent work [2,15], and therefore only brief descriptions are presented below.

For the Eulerian gas phase, the governing equations of mass, momentum, energy, and species mass fraction are solved. They respectively read [16,17]:

$$\frac{\partial(\alpha\rho)}{\partial t} + \nabla \cdot [\alpha\rho\mathbf{u}] = S_{mass} \quad (1)$$

$$\frac{\partial(\alpha\rho\mathbf{u})}{\partial t} + \nabla \cdot [\mathbf{u}(\alpha\rho\mathbf{u})] + \alpha\nabla p + \nabla \cdot (\alpha\mathbf{T}) - \alpha\rho\mathbf{g} = \mathbf{S}_{mom} \quad (2)$$

$$\frac{\partial(\alpha\rho E)}{\partial t} + \nabla \cdot [\mathbf{u}(\alpha\rho E + \alpha p)] + \nabla \cdot [\alpha\mathbf{T} \cdot \mathbf{u}] + \nabla \cdot (\alpha\mathbf{j}) - \alpha\rho\mathbf{g} \cdot \mathbf{u} + p\frac{\partial\alpha}{\partial t} = \alpha\dot{\omega}_T + S_{energy} \quad (3)$$

$$\frac{\partial(\alpha\rho Y_m)}{\partial t} + \nabla \cdot [\mathbf{u}(\alpha\rho Y_m)] + \nabla \cdot (\alpha\mathbf{s_m}) = \alpha\dot{\omega}_m + S_{species,m}, (m = 1, \dots M - 1) \quad (4)$$

In above equations, $\alpha$ is the volume fraction of gas phase, which is calculated from $\alpha = 1 - \sum V_d/V_c$, in which $\sum V_d$ is the total volume of water droplets in a CFD cell and $V_c$ is the cell volume. $t$ is time and $\nabla \cdot (\cdot)$ is the divergence operator. $\rho$ is the gas density, and $\mathbf{u}$ is the gas velocity vector. $p$ is the pressure, calculated from ideal gas equation of state. $\mathbf{T}$ is the viscous stress tensor, whilst $\mathbf{j}$ is the diffusive heat flux. $E$ is the total non-chemical energy. Also, the term $\dot{\omega}_T$ represents the heat release from chemical reactions. $Y_m$ is the mass fraction of $m$-th species, and $M$ is the total species number. $\mathbf{s_m}$ is the species mass flux, and $\dot{\omega}_m$ is the reaction rate of $m$-th species by all $N$ reactions.

For the liquid phase, the Lagrangian method is employed to track the individual droplets. Droplet collisions are neglected because the momentum response time is much shorter than the particle collision time for dilute sprays [17]. Droplet breakup by aerodynamic force from shock impacting is modelled following Pilch and Erdman [18]. Droplet breakup normally occurs above a Weber number $We$ of 12 [19], although it is certainly affected by the Ohnesorge number. Different breakup modes can be modelled based on the total breakup time [18]. It is also assumed that the temperature inside the droplets is uniform due to their small Biot numbers (<0.0013). Gravitational force is not considered due to smallness of the droplets. Therefore, the equations of mass, velocity and temperature of a single droplet read

$$\frac{dm_d}{dt} = -\dot{m}_d, \tag{5}$$

$$\frac{d\mathbf{u}_d}{dt} = \frac{\mathbf{F}_d + \mathbf{F}_p}{m_d}, \tag{6}$$

$$c_{p,d}\frac{dT_d}{dt} = \frac{\dot{Q}_c + \dot{Q}_{lat}}{m_d}, \tag{7}$$

where $m_d$ is the mass of a single droplet. $\dot{m}_d$ is the droplet evaporation rate, estimated from $\dot{m}_d = k_c A_d W_d (c_s - c_g)$ [17]. This model is also used by Duke-Walker et al. [20], demonstrating good accuracies in capturing droplet evaporation after shock passage. $A_d$ is the surface area of a single droplet, $k_c$ the mass transfer coefficient, and $W_d$ the molecular weight of the vapor. $c_s$ and $c_g$ are the vapor mass concentrations at the droplet surface and in the gas phase, respectively.

In Eq. (6), $\mathbf{u}_d$ is the droplet velocity vector. The pressure gradient force $\mathbf{F}_p$ is calculated as $\mathbf{F}_p = -V_d \nabla p$, in which $V_d$ is the volume of a droplet. The total drag force $\mathbf{F}_d$ is $\mathbf{F}_d = (18\mu/\rho_d d_d^2) \cdot (C_d Re_d/24) \cdot m_d(\mathbf{u} - \mathbf{u}_d)$. Here $Re_d$ is the droplet Reynolds number, $C_d$ the drag coefficient predicted by Schiller and Naumann model [21], $\mu$ the gas dynamic viscosity, $\rho_d$ the water material density, $d_d$ the droplet diameter, and $\mathbf{u}$ the gas velocity vector at the droplet location.

In Eq. (7), $c_{p,d}$ is the droplet heat capacity, and $T_d$ is the droplet temperature. The convective heat transfer rate $\dot{Q}_c = h_c A_d (T - T_d)$, where $T$ is the gas temperature at the droplet location and $h_c$ is the convective heat transfer coefficient, estimated with Ranz and Marshall correlations [22]. $\dot{Q}_{lat}$ is the latent heat transfer rate by droplet evaporation. Two-way coupling between gas and droplets are implemented for mass, momentum, and energy exchanges through the source terms of $S_{mass}$, $\mathbf{S}_{mom}$, $S_{energy}$ and $S_{species,m}$ in Eqs. (1)-(4), based on the Particle-source-in-cell (PSI-CELL) method [23].

The gas and liquid phase equations are solved using an OpenFOAM solver *RYrhoCentralFoam* [24–26]. Detailed validations and verifications have been done against experimental and/or theoretical data for a wide range of benchmarking problems [24], including shock capturing, shock-eddy interaction, molecular diffusion, flame-chemistry interactions, single droplet evaporation, and two-phase coupling, and detonation cell size. Therefore, accuracies of the *RYrhoCentralFoam* solver have been quantitatively assessed. Its recent applications can be found from Refs. [2,15,25,27,28].

For the gas phase equations, second-order backward scheme is employed for temporal discretization and the time step is about 7×10$^{-10}$ s. A MUSCL-type scheme [29] with van Leer limiter is used for convective flux calculations in the momentum equations. Total variation diminishing scheme is applied for the convection terms in the energy and species equations. Also, second-order central differencing is applied for all diffusion terms. A detailed mechanism with 13 species and 27 reactions [30] is used for hydrogen combustion. For the liquid phase, Eqs. (5)-(7) are integrated with a Euler implicit method and the right terms are treated with a semi-implicit approach. Details about the numerical method can be found in Refs. [2,24].

## 3. Physical model and numerical implementation

The schematic of the physical problem is shown in Fig. 1. The length ($L$, $x$-direction) and width ($W$) of the computational domain are 0.3 m and 0.025 m, respectively. It includes a driver section ($x = 0 - 0.19$ m, not shown here) and detonation−cloud interaction section ($x = 0.19 - 0.3$ m). One circular cloud is beforehand placed in the second section, to mimic the water sprays generated from a sprinkler to inhibit explosion accident in real applications. The cloud contains many water droplets, and the actual droplet number is determined from the initial droplet diameter $d_d^0$ and droplet concentration $c$. As demonstrated in Fig. 1, the cloud diameter is aligned with $y = 0$ (termed as "centerline" hereafter). The cloud leftmost point (see Fig. 1) always lies at (0.19375 m, 0 m); instead, the cloud center, ($x_c$, 0), is varied when the cloud radius $R$ is changed. Based on our numerical tests, changing the water cloud location (e.g., slightly off-centerline or further downstream) almost does not change its influences on the incident detonation. Therefore, a water cloud is parameterized by its geometry ($R$ or $x_c$) and droplet properties ($d_d^0$ and $c$).

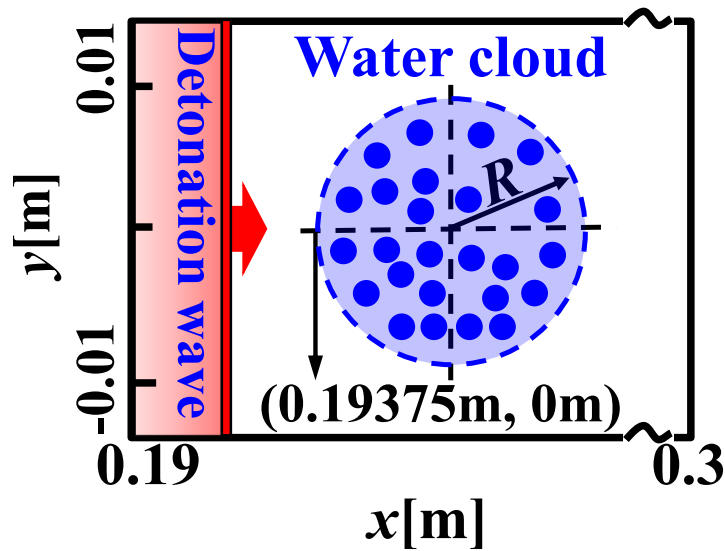

Figure 1: Schematic of detonation-cloud interactions. $R$: cloud radius. Domain and droplet sizes not to scale.

Initially ($t= 0$), the entire domain is filled with stoichiometric $H_2$/air mixture, with temperature and pressure being $T_0 = 300$ K and $p_0 = 50$ kPa, respectively. The detonation is initiated by three vertically placed hot spots (2,000 K and 5 MPa) near the left end ($x$ = 0 m). Inside the cloud, mono-sized water droplets (diameter $d_d^0$ = 1 − 15 $\mu m$, and concentration $c$ = 0.105 − 1.68 kg/m³) are uniformly distributed. They are static ($\mathbf{u}_d^0$ = 0) before the detonation wave arrives. Their initial temperature, material density and heat capacity are 300 K, 997 kg/m³ and 4,187 J/kg·K, respectively. The radius of the water cloud, $R$, varies from 0.00625 to 0.0125 m, corresponding to 25%-50% of the domain width $W$.

The upper and lower boundaries of the domain in Fig. 1 are periodic. For the left boundary ($x$ = 0), the non-reflective condition is enforced for the pressure, while the zero-gradient condition applied for other quantities. For the boundary at $x$ = 0.3 m, zero gradient is assumed for all variables [31].

The domain in Fig. 1 is discretized with uniform Cartesian cells of 20×20 μm². The half-reaction length in the ZND structure of stoichiometric H₂/air detonation under the current investigated conditions is approximately 382 $\mu m$. Therefore, over 19 cells can be expected in the induction zone in our simulations, because of the thickened hydrodynamic structure due to droplet breakup, heating and vaporization [2,12]. Besides, mesh sensitivity is tested, showing that the detonation cellular structure is almost not affected when the mesh size is 10×10 $\mu m^2$ (see Fig. S1 of the supplementary document). Moreover, we assess the relative errors of droplet evaporation timescale [32] with current Eulerian mesh resolution and Lagrangian droplet size, and find that they are generally below 10% due to the strong Pelect number effects in detonated flows (see Fig. S2 of the supplementary document). This is consistent with our previous finding from 1D gas-droplet detonation modelling [15] and hence confirms the accuracy of droplet evaporation predictions in our simulations.

## 4. Results and discussion

### 4.1 Water droplet size and concentration effects

The water droplet size and concentration effects will be first studied in this section, whilst the cloud size effects will be discussed in Section 4.3. Figure 2 shows the evolutions of peak pressure trajectories in detonation-cloud interactions with different droplet concentrations $c$. The initial droplet diameter and cloud radius are fixed to be 2.5 μm and $R = 0.25W$, respectively. The cloud-free ($c = 0$) results are shown in Fig. 2(a) for comparison. Evident from Fig. 2 is that the detonation cellular structures change, with different degrees, after the DWs cross the water clouds (initial position marked as blue circles, and their movement will be discussed in Section 4.4). For low droplet concentration, e.g., $c = 0.105$ kg/m³ in Fig. 2(b), the averaged cell width behind the water cloud is slightly increased, indicating the enhanced frontal instability in the post-cloud area. Moreover, the peak pressures inside the cloud are reduced, indicating the reduced chemical reactivity at the triple points of the refracted DW due to droplet heating and evaporation. Nonetheless, the cell size and regularity are quickly restored when $x > 0.23$ m. We term this mode as *perturbed propagation*.

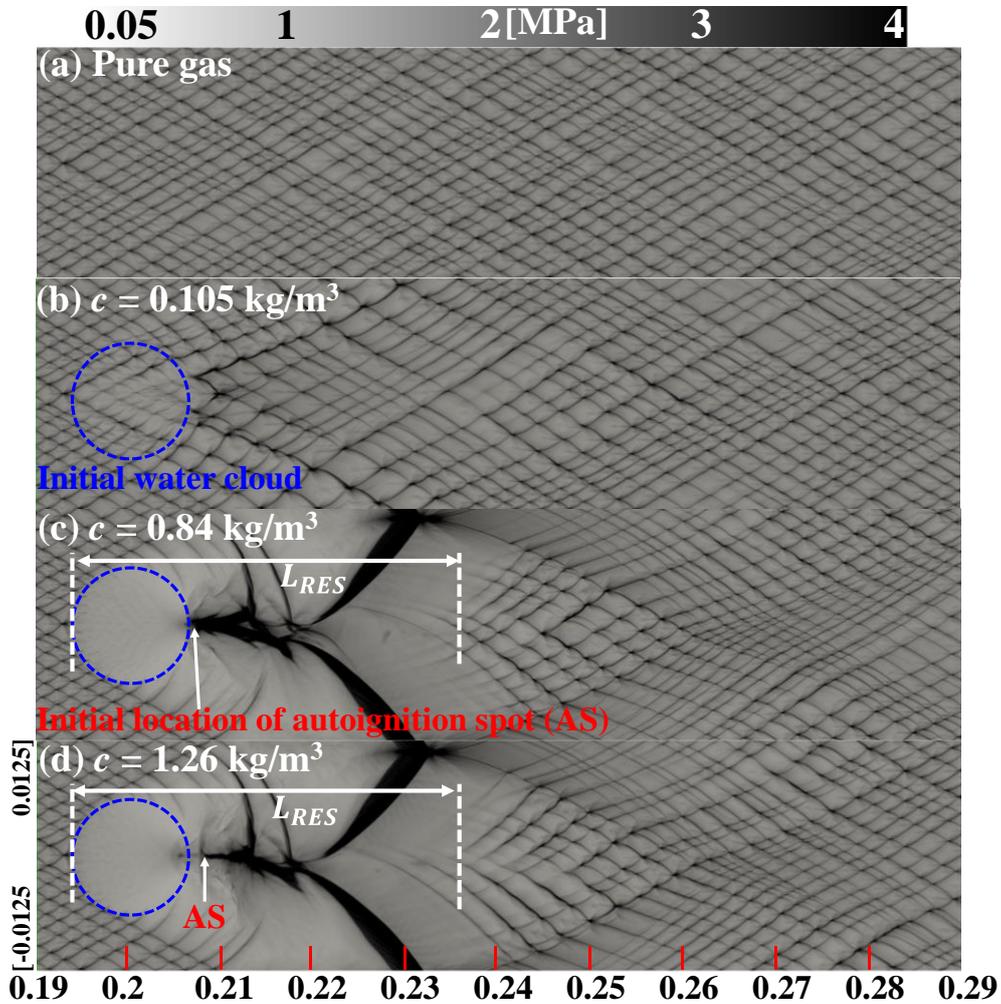

Figure 2 Trajectories of peak pressure with various droplet concentrations. $d_d^0 = 2.5$ $\mu m$ and $R = 0.25W$. Axis label unit: m.

When $c \geq 0.84$ kg/m$^3$, maximum pressures are immediately reduced to below 1 MPa in the cloud and the cellular structures are not observable (see Figs. 2c and 2d). This means that the chemical reactions at the triple points are significantly weakened because of the high-concentration water mists. Actually, the refracted DWs in the cloud experience localized extinctions, featured by decoupled shock front (SF) and reaction front (RF) at most locations of the detonation front. Downstream the southern/northern poles, the peak pressures from the diffracted DWs are also significantly reduced, which may indicate detonation failure after diffraction. Behind the cloud, the peak pressure trajectories become blurred.

A horizontal black strip with high overpressure extends from the downstream pole, which corresponds to the evolution of an igniting hot spot (labelled with AS in Fig. 2c) due to shock compression. In Figs. 2(c) and 2(d), at about $x = 0.22$ m, the strip bifurcates into two thick trajectories, which are the upper and lower transverse detonation connected with a Mach stem generated from a re-initiation process. At $x$ = 0.24 m, cellular structures appear again. The mean cell widths are close (about 1.9 mm) in Figs. 2(c) and 2(d), similar to that (1.7 mm) before the DW interacts with the water cloud. The distance between the cloud upstream pole and minimum $x$ coordinate with clear cellular structure can be defined as DW restoration distance, which is $L_{RES} \approx$ 41 mm in Figs. 2(c) and 2(d). We term this process as *leeward re-detonation*. The detailed transients and underlying mechanism will be discussed in Section 4.2. From this mode, the useful implication for practical hydrogen detonation mitigation implementation is that the overpressure only in some areas, i.e., immediately downstream of the cloud, can be reduced. Nonetheless, the detonation wave cannot be ultimately quenched by a single cloud of the studied properties in Fig. 2.

Figure 3 presents the peak pressure trajectories with varying droplet diameters, $d_d^0 = 1, 5$, and 15 μm. Here $c$ = 0.84 kg/m$^3$ and $R = 0.25W$. Note that the results of 2.5 μm are already shown in Fig. 2(c). For small droplets (e.g., 1 and 5 μm), the DWs experience leeward re-detonation process and the evolutions of cellular structures are like what have been discussed in Fig. 2(c). Their DW restoration distances are respectively about 43 and 41 mm, annotated in Figs. 3(a) and 3(b). However, for large droplets in Fig. 3(c), DW extinction does not occur because of slower interphase exchange rates for mass and energy; instead, only perturbed DW propagation occurs. Specifically, the detonation cells become more irregular, and the averaged cell width is increased, corresponding to more unstable front after it crosses the cloud.

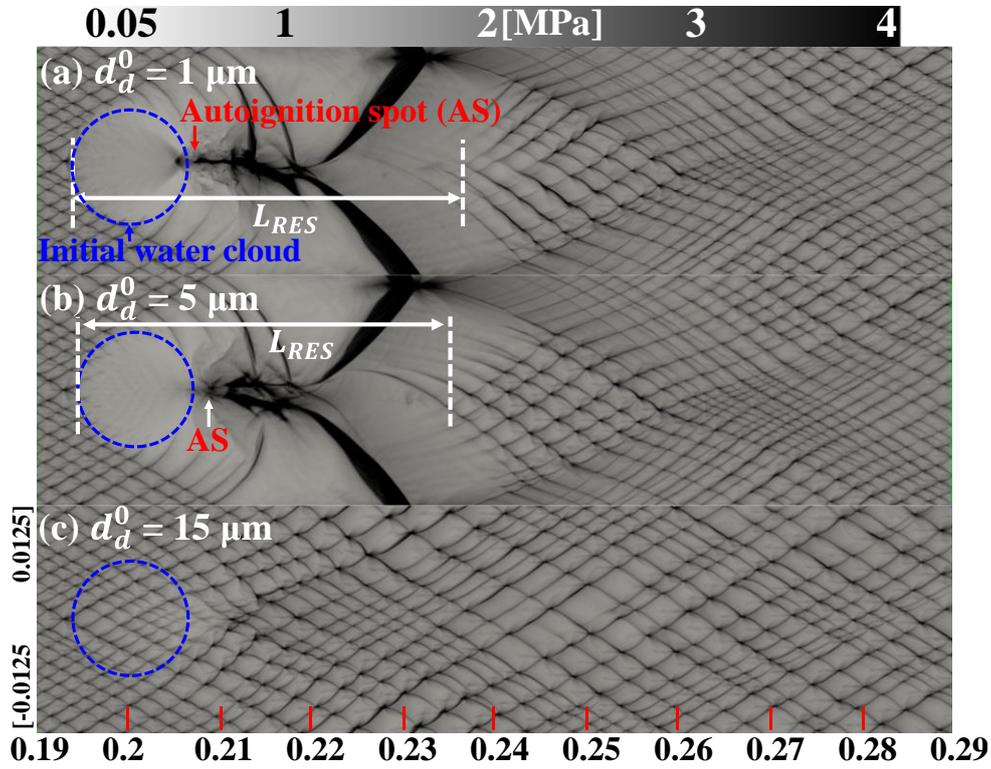

Figure 3 Trajectories of peak pressure with various droplet diameters. $c = 0.84$ kg/m$^3$ and $R = 0.25W$. Axis label unit: m.

The speeds of the lead shocks when they propagate inside the cloud with various droplet concentrations and diameters are shown in Fig. 4. They are scaled by the C-J speed of droplet-free H$_2$/air mixture, i.e., $D_{CJ} = 1,961$ m/s. Cloud-free case (black line) is included for comparison. Generally, the water cloud exhibits significant effects on the shock propagation. The shock speed within the cloud is lower than $D_{CJ}$. It decreases when the droplet concentration (diameter) becomes larger (smaller). This tendency is associated with increased heat and momentum transferred from the background gas to the dispersed phase. For those extinction / re-detonation scenarios, e.g., $c = 1.26$ kg/m$^3$ in Fig. 4(a) or $d_d^0 = 1$ μm in Fig. 4(b), the shock speed is even as low as 0.4-0.5$D_{CJ}$.

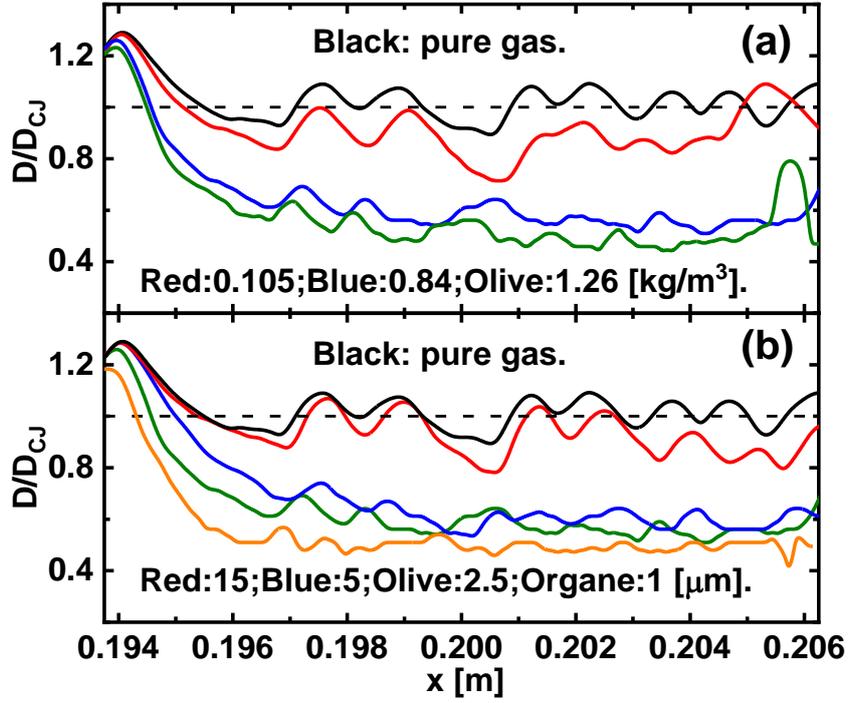

Figure 4 Lead shock speed in the cloud as a function of droplet concentration (a) and diameter (b). $R = 0.25W$. The droplet diameter in Fig. 4(a) is 2.5 $\mu m$, and the droplet concentration in Fig. 4(b) is 0.84 kg/m$^3$.

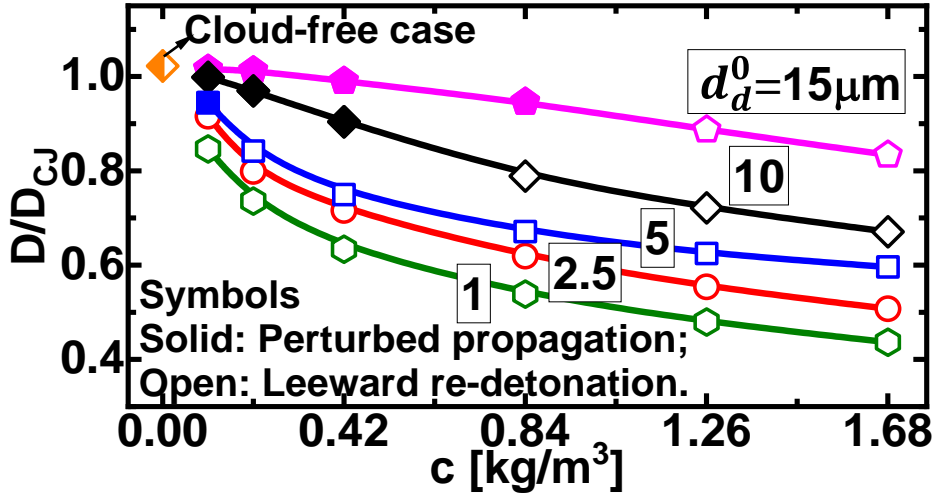

Figure 5 Average shock speed for crossing water cloud as a function of droplet concentration and diameter. $R = 0.25W$.

Detonation-cloud interactions are further simulated considering a broader range of droplet concentration and diameter. We compare the average lead SF propagation speed in the cloud, $D$, which is calculated from the cloud diameter ($2R$) divided by the time duration with which the shock respectively crosses the upstream and downstream poles along the centerline ($y = 0$). Plotted in Fig. 5 are the lead SF speed $D$ when the cloud size is $R = 0.25W$. Due to relatively small cloud size, detonation full extinction (i.e., SF-RF decoupling) does not occur in all simulated cases; instead, only perturbed propagation and leeward re-detonation modes are observed, which are distinguished in Fig. 5 with solid and open symbols, respectively. Note that the speed in the cloud-free case is calculated based on the time for the DW to pass the upstream and downstream poles of the cloud, although the cloud is not placed. One can see that perturbed detonation propagation only occurs when the droplet size is large and water concentration is small. Generally, $D$ is below the DW propagation speed in the cloud-free case. For a constant droplet concertation (e.g., 0.84 kg/m³), $D$ decreases when the droplet size becomes smaller. This is because the smaller diameter corresponds to a larger specific surface area, and therefore faster interphase exchange of mass, momentum, and energy. Moreover, when the droplet diameter is fixed (e.g., 15 $\mu m$), $D$ decreases with $c$. This is caused by the enhanced influences of fine droplets on the DW due to increased droplet concentration. Meanwhile, the dependence of SF speed on the droplet concentration becomes more pronounced when the droplet size is smaller.

**4.2 Leeward re-detonation**

The leeward re-detonation will be further interpreted based on the case of $c$ = 0.84 kg/m³, $d_d^0$ = 2.5 $\mu m$, and $R$ = 0.25 $W$. This corresponds to Fig. 2(c). Following Ling et al. [33] and Sommerfeld [34], the heterogeneous mixture of gas and droplets inside the cloud can be deemed a "equivalent fluid". The initial density and sound speed of this equivalent fluid are 1.26 kg/m³ and 210 m/s, respectively. The specific acoustic impedance (product of density and sound speed) is hence about 264 kg/m²/s, slightly higher than that of the surrounding gas (171 kg/m²/s). Due to the impedance mismatch at the cloud boundary, the cloud acoustically behaves like a "converging lens", leading to refraction and diffraction of the incident DW [35,36]. However, since the impedance difference is still limited due to dilute droplet concentrations, no reflected shocks from the upstream boundary are observed.

Figure 6 shows the time sequence of gas temperature during a leeward re-detonation process. An animation of this event is available in the supplementary material. Specifically, at 98 $\mu s$ in Fig. 6(a), the temperature behind the refracted DW inside the cloud (marked as inner DW) is reduced to around 2,150 K, much lower than that before the DW arrives at the cloud. At 102 $\mu s$, the RF and SF inside the cloud decouple. The RF is featured by further reduced temperature (~1,600 K) and HRR (~$1\times10^{12}$ J/m$^3$/s). However, beyond the cloud, the detonation wave (termed as outer DW) remains. The outer DW starts to diffract after passing the southern/northern poles of the cloud and propagates relatively fast compared to the quenched one in the cloud. Due to the small impedance distinction between the two fluids, one can conjecture that the refracted wave deceleration is largely affected by the reduced chemistry due to heat absorption and/or vapor dilution by the evaporating cloud. Moreover, the outer (diffracted, burning) and inner (refracted, quenched) sections are connected at the cloud boundary (dashed line). Meanwhile, the outer DW becomes slightly curved near the cloud boundary because of wave diffraction. From 104 to 107 $\mu s$, the outer DW becomes convex with respect to the fresh gas, and the upper / lower RFs move gradually closer. Meanwhile, decoupling of SF and RF also occur for the outer DW near the cloud. This is also observed in the detonation diffraction studies [37–40], and it is attributed to the interactions between the detonation front and expansion waves resulting from the diverging geometry of the cloud downstream boundary.

In Fig. 6(e), shock focusing occurs near the centerline: the diffracted SFs degraded from the upper and lower parts of the outer DW are superimposed near the centerline. This leads to high overpressure (> 3 MPa) and temperature (> 1,600 K) at the point S1. This area is compressed again by the late refracted wave and hence therefore the local thermochemical conditions are further elevated towards the threshold of autoignition. It should be mentioned that the sequence of diffracted SF focusing and refracted shock arrival largely depends on the wave speed difference inside and outside the cloud [41,42]; their more pronounced difference may lead to an implosion region enclosed with the refracted and diffracted shocks [43]. Meanwhile, two peninsula-shaped RFs from the upper and lower DWs are approaching each other near the focused point, which also provides a favorable environment full of energetic radicals to promote the local chemical reactions. At 108.5 $\mu s$, an isolated autoigniting spot (AS) arises there, which generates higher pressure and temperature, indicating the onset of thermal runaway in an isochoric combustion mode. As shown in Fig. 6(f), the AS lies at the right of the cloud boundary and hence autoignition proceeds at the cloud leeward side, i.e., in a gas-only mixture. Subsequently, AS quickly develops into detonation waves at 111 $\mu s$ in Fig. 6(h). Their transverse propagation consumes the shocked gas between the lead SF and RF. It also overtakes the lead SF and an overdriven Mach stem is formed in Fig. 6(i). The upward- and downward-running transverse DWs generate the thick bifurcated trajectories as exhibited in Fig. 2(c).

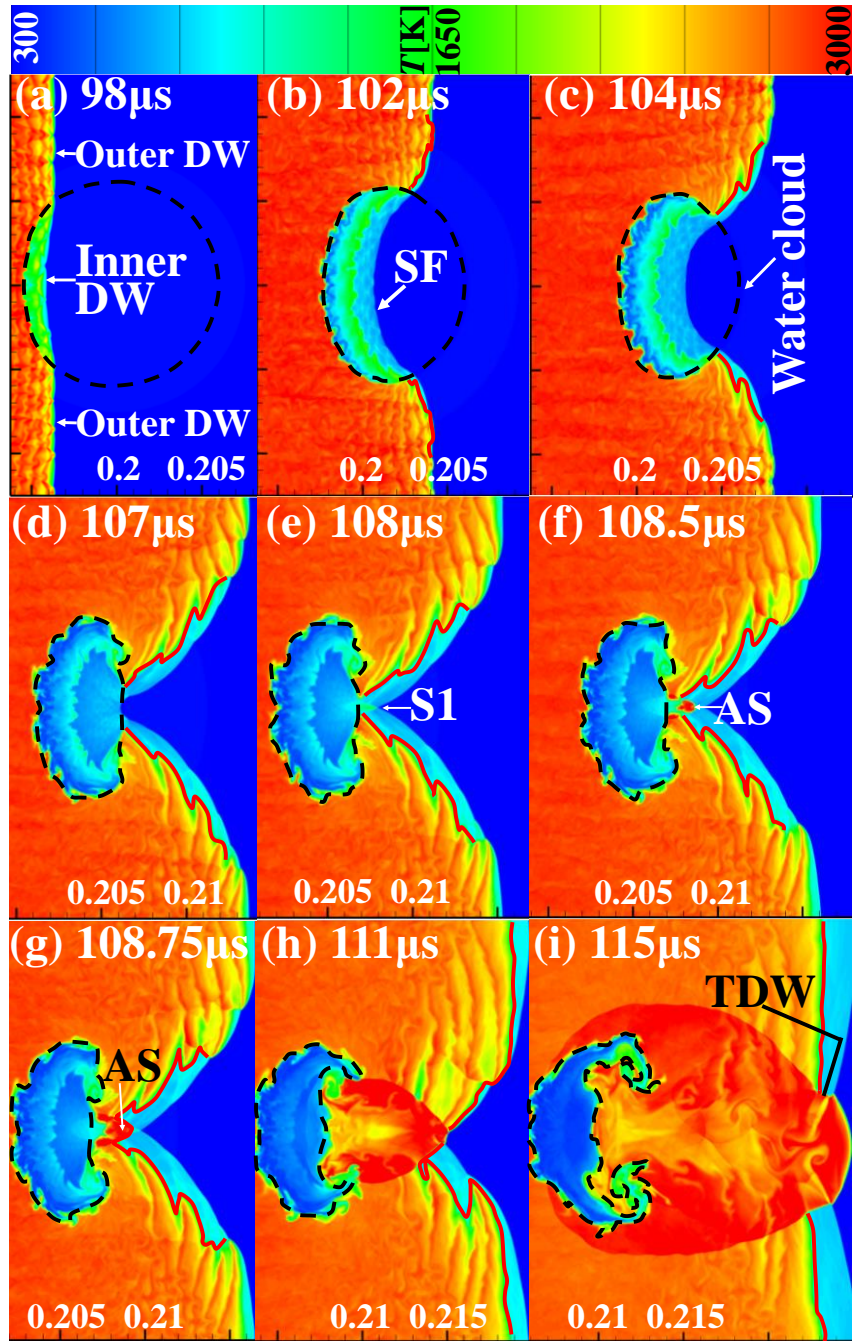

Figure 6 Evolutions of gas temperature in a leeward re-detonation process. $c$ = 0.84 kg/m³, $d_d^0$= 2.5 $\mu m$, $R$ = 0.25$W$. Axis label unit: m. Dashed lines: cloud boundary; RF/SF: reaction/shock fronts; AS: auto-ignition spot; TDW: transverse detonation wave.

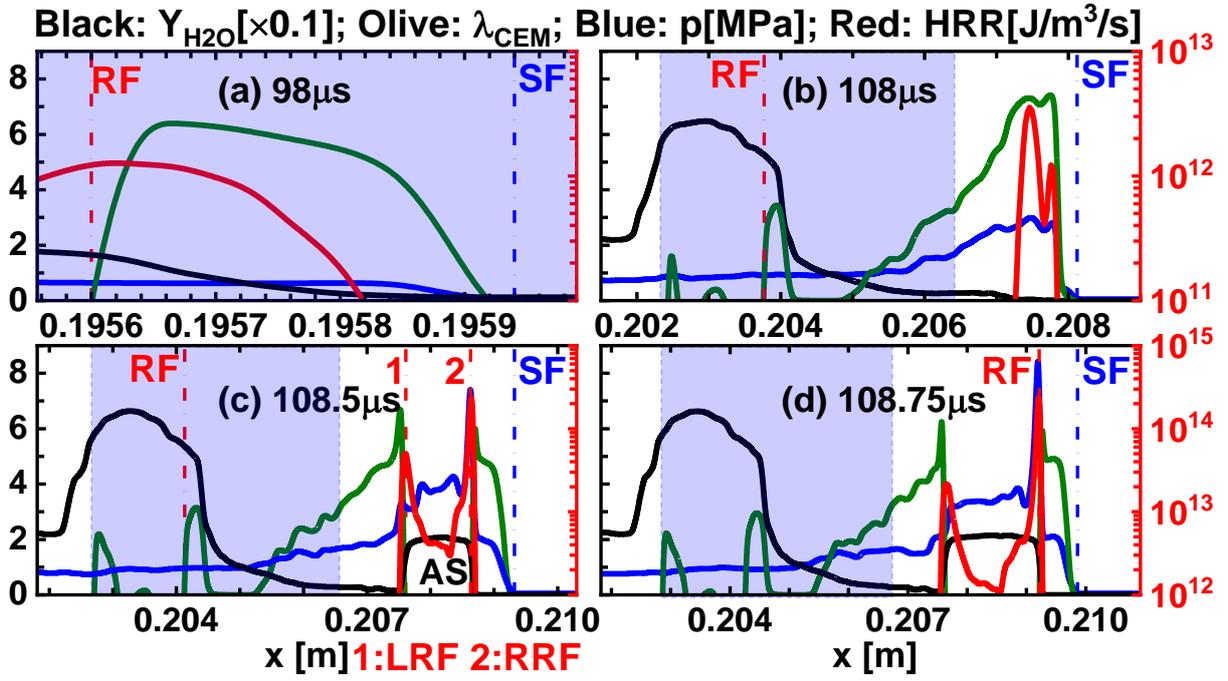

Figure 7 Evolutions of $\lambda_{CEM}$, pressure, HRR, and water vapor mass fraction around the autoigniting spot. $c = 0.84$ kg/m$^3$, $d_d^0 = 2.5$ μm, and $R = 0.25W$. Shaded area: water cloud.

To further reveal the chemistry of the leeward re-detonation process, chemical explosive mode analysis [44–46] is performed to extract the characteristic chemical information of the hydrogen detonation. Figure 7 shows the spatial profiles of pressure, HRR, and eigen values $\lambda_e$ of the chemical Jacobian along the centerline at four instants. For clear illustration, the signed exponent of the eigenvalue $\lambda_e$ of chemical explosive mode is plotted, i.e., $\lambda_{CEM} = sign[Re(\lambda_e)] \cdot log_{10}[1 + |Re(\lambda_e)|]$. Positive (negative) $\lambda_e$ indicates the propensity of chemical explosion (non-reactive or burned mixtures). In Fig. 7(a), the RF and SF lie in the water cloud (shaded area). Despite detonation extinction, the mixture in the induction zone between RF and SF still has explosion propensity with high $\lambda_{CEM}$. Meanwhile, the pressure behind the SF is slightly increased because of shock compression. The water vapor mass fraction gradually increases towards the RF. At 108 μs in Fig. 7(b), $\lambda_{CEM}$ peaks behind the SF due to shock focusing, evidenced by the elevated pressure, HRR peaks and high $\lambda_{CEM}$ at the same location. At $x = 0.204$-$0.205$ m, the mixture reactivity is low (smaller $\lambda_{CEM}$) because of droplet evaporation and/or interphase heat transfer inside the cloud. This can also be confirmed by finite water vapor mass fraction $Y_{H2O}$ (cf. Fig. S3 of the supplementary document) ahead of the RF. In Fig. 7(c), the autoigniting spot (i.e., AS in Fig. 6f) is generated, corresponding to locally negative $\lambda_{CEM}$ and two RFs with high HRR (> 4×10$^{13}$ J/m$^3$/s) at $x$ = 0.2076 and 0.2086 m. For the same reason, pressure is also increased near the AS location in Figs. 7(c) and 7(d). Moreover, at 108.75 μs, the right RF is coupled with the SF, generating a new detonation wave (i.e., evolving into the Mach stem in Fig. 6i).

Reaction initiation due to focusing of the oppositely propagating diffracted waves is also observed in other re-detonation cases from our simulations, as indicated in Fig. 5. Although shock focusing has been discussed in previous studies on shock-bubble interactions, e.g., in [42], nonetheless, its consequence in a flammable gas (e.g., hydrogen) has not been studied before. In some previous experimental and computational studies on the interactions between shock and H$_2$/O$_2$ bubble surrounded by N$_2$ [43,47–49], ignition induced by shock focusing only occurs inside the reactive bubble when the pressure is relatively low. For more reactive conditions, detonation development is found, mainly from direct initiation by strong shock compression. Differently, from our cases, AS always occurs along the centerline behind the water cloud, which may be because of the existence of the evaporating water in the cloud. We compile the AS locations from all our re-detonation cases in Fig. S5 of the supplementary document. It is interesting to find that the AS locations ($x$ coordinate) range from 2.1$R$ to 3.3$R$, relative to the upstream pole of the cloud. This corresponds to external shock focusing [42], which is essentially determined by the differentiated timescales for diffracted and refracted shocks [41].

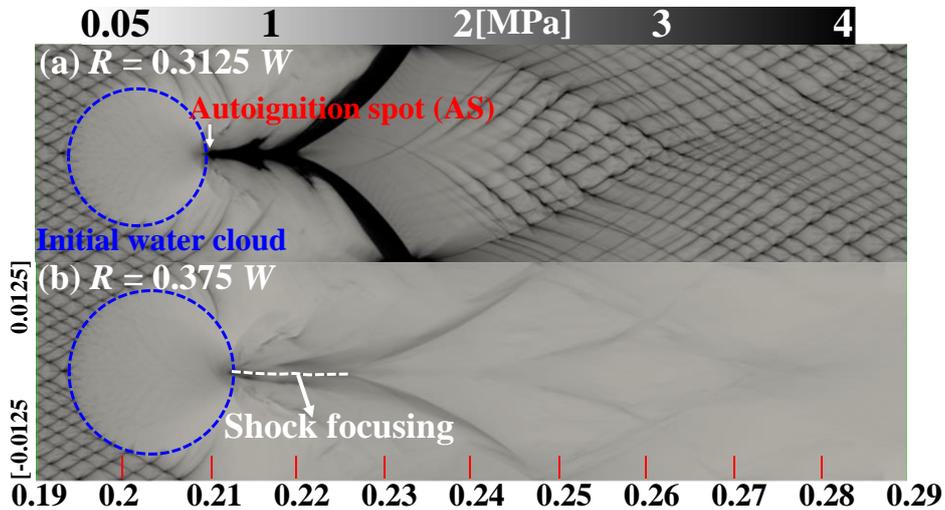

Figure 8 Numerical soot foils change with various cloud sizes. $c$ = 0.84 kg/m$^3$ and $d_d^0$= 2.5 $\mu m$. Axis label unit: m.

## 4.3 Cloud size effects and detonation extinction

Up to this point, the circular cloud radius is fixed to $R = 0.25W$. The effects of cloud size on the incident detonation wave will be studied in this section, through the peak pressure trajectories in Fig. 8. Two additional cloud radii, $R = 0.33W$ and $0.375W$, are considered with $c = 0.84$ kg/m$^3$ and $d_d^0 =$ 2.5 $\mu m$. For the cloud size in Fig. 8(a), the above-mentioned extinction / re-detonation process in Fig. 2(c) with $R = 0.25 W$ is observed. However, in the larger cloud in Fig. 8(b), the incident DW is quenched after crossing it, characterized by quickly fading peak pressure trajectories, and no re-initiation is observed. The transient of DW extinction in Fig. 8(b) is visualized in Fig. 9 and through an animation in the supplementary material. The detonation frontal structures in Figs. 9(a)-9(d) share the similar evolutions with Fig. 6. One can see that both refracted and diffracted DWs are quenched, and shock focusing occurs at S1 in Fig. 9(e). However, in this scenario, no autoigniting spot is observed, and the distance between the lead SF and RF is gradually lengthened, as shown in Fig. 9(f).

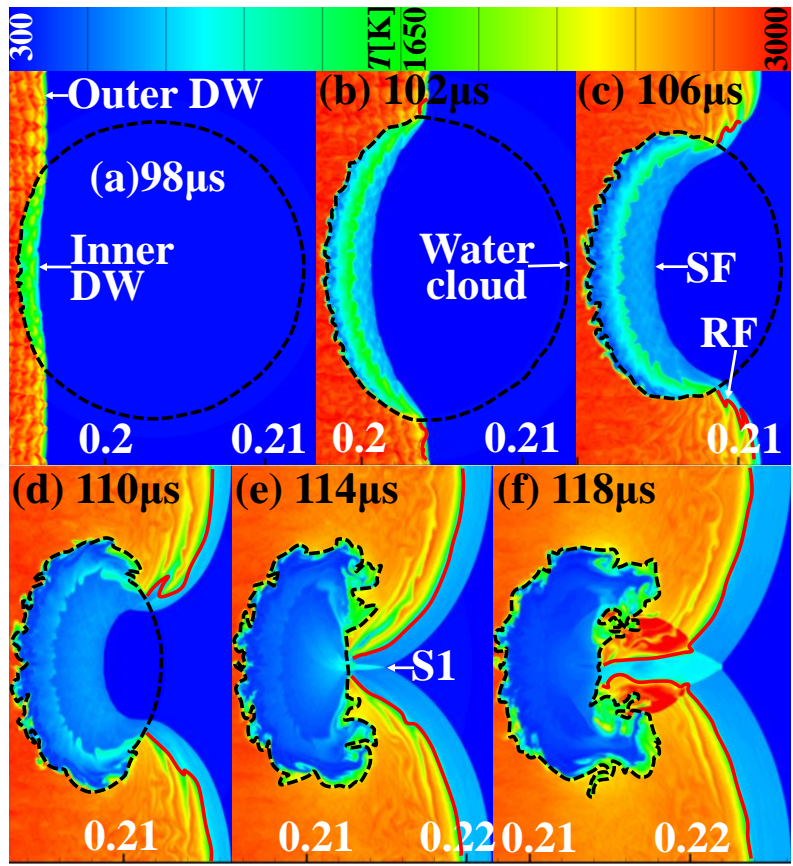

Figure 9 Time history of gas temperature in a quenched detonation. $c = 0.84$ kg/m$^3$, $d_d^0 = 2.5$ $\mu m$, and $R = 0.375 W$. S1: shock focusing location. Axis label unit: m.

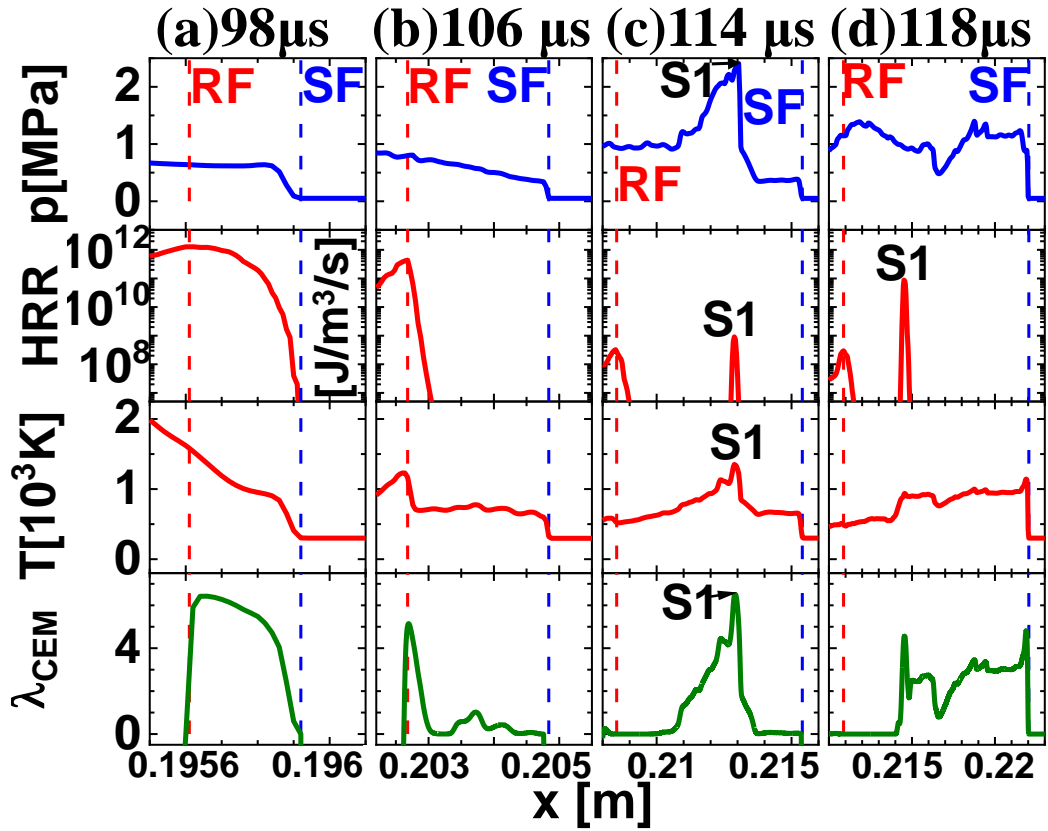

Figure 10 Evolutions of thermochemical structures along the centerline between the RF and SF. $c$ = 0.84 kg/m$^3$, $d_d^0$= 2.5 $\mu m$, and $R$ = 0.375$W$.

To interpret the reaction initiation failure in shock superimposition in Fig. 9, time history of pressure, HRR, gas temperature, and $\lambda_{CEM}$ are plotted in Fig. 10. They are extracted along the centerline in Fig. 9. The peak values of pressure, temperature, HRR and $\lambda_{CEM}$ at S1 in Fig. 10(c) is lower than the counterpart of S1 in Fig. 6(b). The differentiated thermochemical conditions in the two autoignition spots can be found from Table S1 of the supplementary document. Specifically, peak pressure (2.4 MPa) and temperature (1,344 K) from this case are much lower than those in the successful one (3 MPa and 1,602 K, respectively). This, therefore, cannot induce an AS in Fig. 10(c), although the shocked gas mixture is highly chemically explosive around S1 ($\lambda_{CEM}$ = 6.5). This can be further confirmed by the reduction of pressure, temperature and $\lambda_{CEM}$ around S1 in Fig. 10(d), and almost unconsumed H$_2$/O$_2$ around S1 (cf. Fig. S4 of the supplementary document).

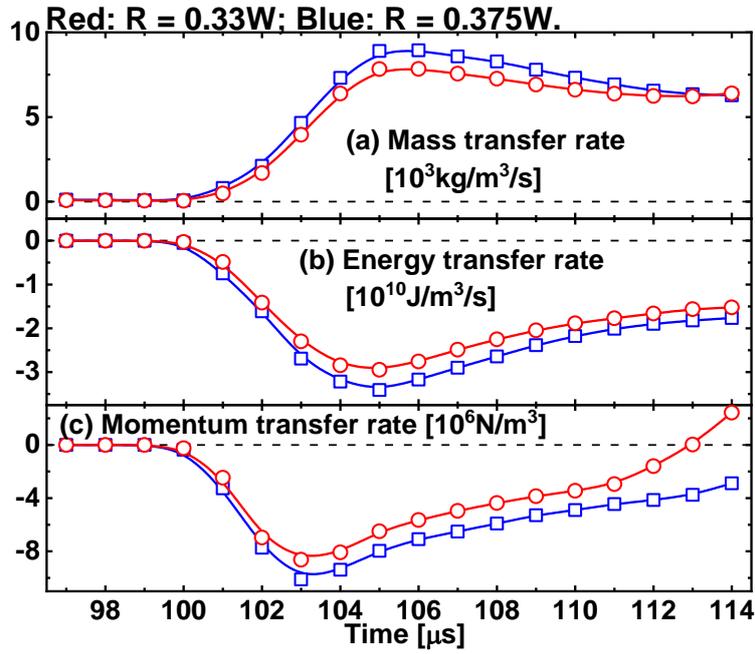

Figure 11 Interphase transfer rates in detonation transmission and failure cases.

The interphase transfer rates of mass, energy and momentum of detonation transmission and failure transients from the two cases in Figs. 9 are compared in Fig. 11. They are averaged from the whole domain ($x = 0 - 0.3$ m). Positive transfer rate indicates the transfer proceeds from droplet to gas phase. Note that the DW enters and leaves the water cloud at 97 and 112/114 (red/blue lines) μs, respectively. Results in Fig. 11 mainly reflect the interactions between the refracted DW and cloud since the diffracted one only propagates in the gaseous medium. The results from two cases bear some resemblance. For instance, all transfer rate magnitudes first increase, and then gradually decrease. This is because the droplet number in the shocked area reaches the peak at around 103 μs, as the transmitted shock reaches the downstream pole. However, it is demonstrated that larger water cloud (i.e., $0.375W$) generally has higher transfer rates, which leads to lower chemical reactivity as discussed in Fig. 10.

Figure 12 demonstrates the critical cloud size of DW propagation and extinction. The droplet diameter is fixed to be 2.5 μm, but various droplet concentrations ($c$=0.105─1.68 kg/m$^3$) are considered. It should be clarified that here "extinction" indicates the overall outcome of the incident detonation, instead of the refracted or diffracted one. It is observed that the critical water cloud diameter, $2R$, generally decreases when the droplet concentration is increased. Below or left to this curve, perturbed propagation (only observed for relatively small $c$) or re-detonation occurs, whilst above it the incident detonation wave is fully quenched behind the water cloud. When $c$ is larger than 0.84 kg/m$^3$, the critical cloud size is negligibly influenced by the droplet concentration. This implies that beyond some critical concentration the performance of water cloud in inhibiting detonation may be limited by droplet size, which directly affects droplet relaxation process and evaporation. This is consistent with the finding from our previous work [15].

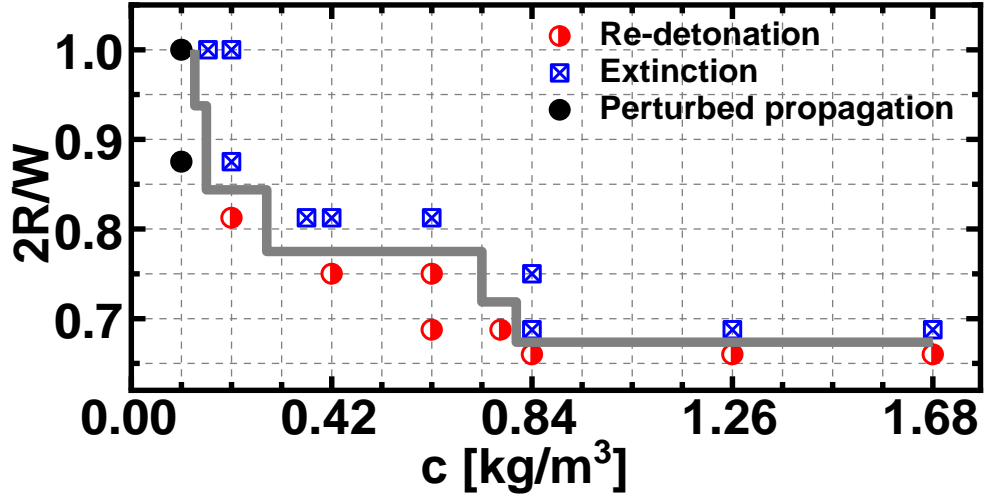

Figure 12 Diagram of DW propagation and extinction. $d_d^0$= 2.5 $\mu m$.

**4.4 Two-phase fluid interfacial instability**

Two-phase fluid interfacial instability subject to detonation wave impacting will be discussed in this section. The cloud boundary is deemed as an interface separating the background gas and two-phase gas-droplet mixture. Following Richtmyer–Meshkov instability (RMI) theory [50], an effective Atwood number can be used to quantify the density difference across this interface [20], i.e.,

$$A_e \equiv (\rho_{e2} - \rho_{e1})/(\rho_{e2} + \rho_{e1}), \tag{8}$$

where $\rho_{e1}$ and $\rho_{e2}$ are the effective densities of the background gas and the gas-liquid mixture in the water cloud, respectively. They can be calculated from gas phase volume fraction $\alpha$, gas density $\rho$, and droplet material density $\rho_d$, i.e., $\rho_{e2} = (1-\alpha)\rho_d + \alpha\rho$ [17].

Figure 13 shows that the effective Atwood number $A_e$ in our simulations cases. It is seen that $A_e$ monotonically increases with the droplet concentration, i.e., 0.11 to 0.67. Note that the droplet diameter does not influence $A_e$ for a fixed concentration. Based on our results, there are two types of cloud morphology evolutions after being struck by the incident detonation wave: shrinking clouds with tails (i.e., relatively larger $A_e$, i.e., 0.33-0.67) and without tails (smaller $A_e$), as indicated in Fig. 13. Their detailed evolutions and underpinning mechanisms will be discussed below.

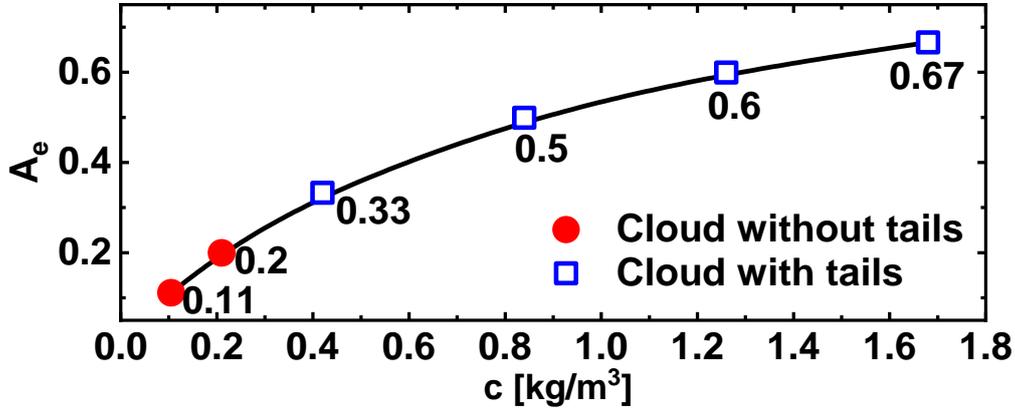

Figure 13 Effective Atwood number as a function of droplet concentration.

**4.4.1 Mechanism of cloud evolution**

We first discuss the case of $A_e = 0.5$, based on the time sequence of Lagrangian water droplet distributions in Fig. 14. The conditions are $c = 0.84$ kg/m³, $d_d^0 = 2.5$ μm, and $R = 0.25W$. Note that the corresponding gas dynamics have been studied in Fig. 6. At 96 μs, the cloud is intact: the droplet size is almost equal to the initial value, indicating limited evaporation before the DW arrives. At 100 μs, the refracted SW (dashed line in Fig. 14b) encroaches the left part of the cloud. The droplets behind it start to break up, resulting in quickly reduced droplet size (< 1.5 μm) and hence strong evaporation. At 100-106 μs, the left cloud boundary, carrying many disintegrated droplets, gradually moves with the refracted shock, which leads to a shrinking cloud.

At 108 μs, the shock just leaves the cloud and most droplets are highly disintegrated, evidenced by the reduced droplet size and increased droplet number (by about 60%, from our results). The Weber numbers, $We$, of these droplets range from 40 to 70, and therefore bag-and-stamen breakup mechanism is dominant [18]. The $We$ is slightly higher than those (5-33, hence bag breakup mechanism) from the shock-water cloud experiments by Middlebrooks et al. [51]. This is reasonable since weaker shocks ($Ma = 1.66$) are used in their tests. Closer inspection of Fig. 14(f) shows that some jets of fine droplets (below 1 μm) appear near the southern/northern poles of the cloud. This phenomenon is caused by the fast response of the fine droplets to the local aerodynamics, i.e., vortex shedding due to the impulsive shock acceleration of a perturbed interface between multiphase fluids [20,52]. The vortex evolutions will be further interpreted in Fig. 15. Meanwhile, in Fig. 14(f), the evolutions of vortices further lead to two tails of fine droplets extending from the cloud. At the subsequent instants, e.g., 112-130 μs, these tails are continuously stretched relative to the cloud, until the droplets along them are vaporized at 154 μs.

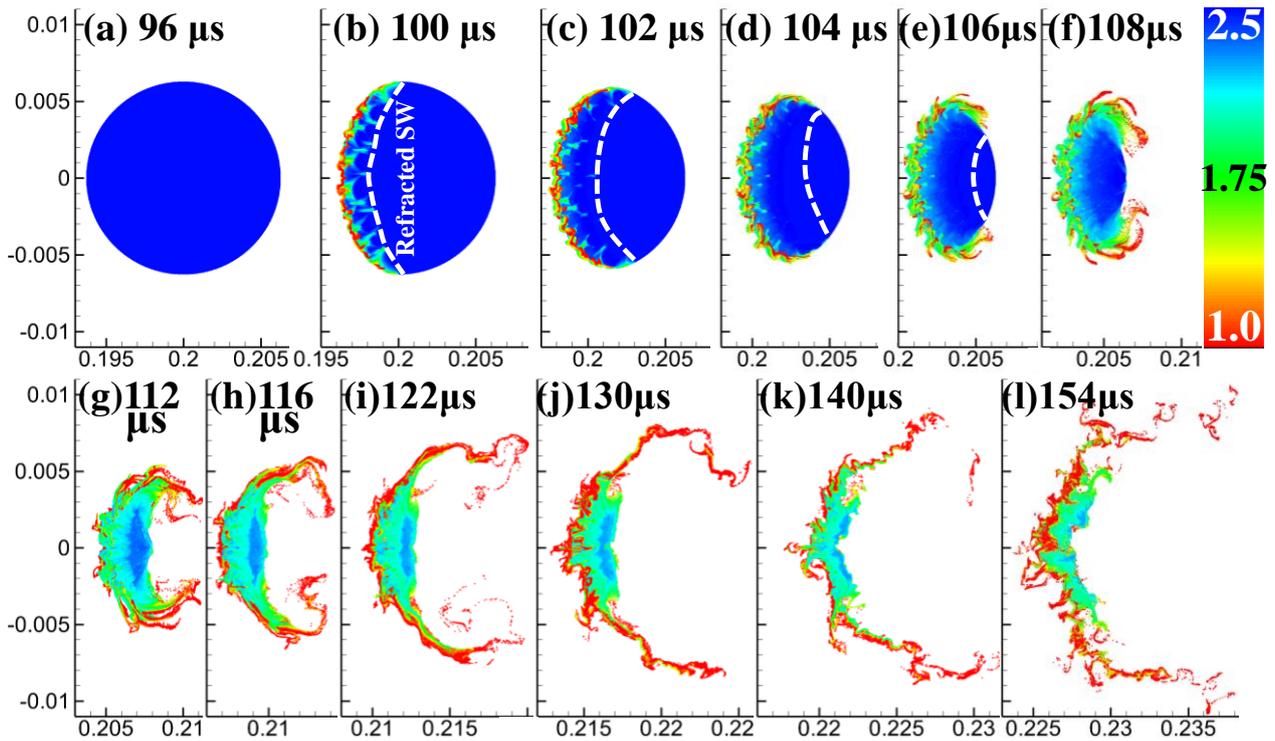

Figure 14 Evolutions of Lagrangian droplets colored by diameter (in μm). Dashed curve: refracted wave. $A_e = 0.5$, $c = 0.84$ kg/m$^3$, $d_d^0 = 2.5$ μm, and $R = 0.25W$. Axis label unit: m.

To better understand the vortex development near the cloud, Fig. 15 shows the evolutions of out-of-plane component of vorticity, $\omega_z$, at four selected instants of Fig. 14. Positive (negative) $\omega_z$ means that the local fluid rotation is counter-clockwise (clockwise). In the shocked area behind the SW, lots of small eddies can be seen. Nonetheless, inside the water cloud, the magnitude of $\omega_z$ is significantly reduced, to approximately zero. This results from the local momentum exchanges between the two phases, which to some degree stabilizes the background gas [51]. Along the cloud boundary (i.e., fluid interface), connected zones with high magnitudes of $\omega_z$ can be seen, i.e., dominantly positive (negative) $\omega_z$ at the lower (upper) part. At 108 $\mu s$, a pair of well-defined counter-rotating eddies starts near the northern and southern poles, which become stronger at 116 $\mu s$. Based on Fig. 14(h), the two droplet-carrying tails roughly follow the counterrotating vortices. At 154 $\mu s$, although a pair of vortices are still observable, however, the vorticity becomes weaker and more localized, compared with earlier instants.

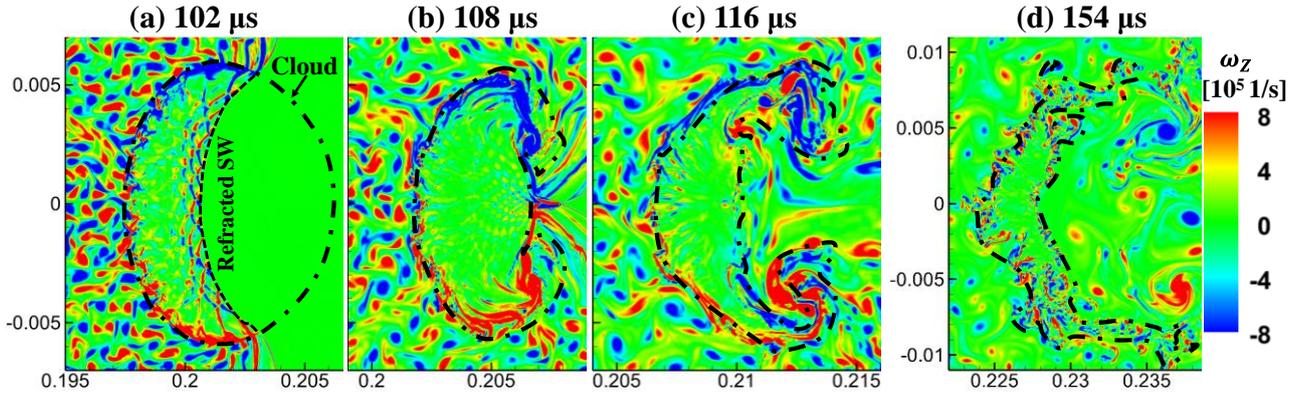

Figure 15 Time sequence of *z*-component vorticity. Dashed line: refracted wave. Dash-dotted line: cloud boundary. Axis label unit: m.

Evolutions of the density and pressure gradient magnitudes ($|\nabla \rho|$ and $|\nabla p|$) of Fig. 15 are shown in Fig. 16. In Fig. 16(a), in the cloud, $|\nabla \rho|$ is zero ahead of the SF. This is because no water vapor is added in the premixed hydrogen/air gas before the shock arrives. However, along the SF and RF, $|\nabla \rho|$ is considerable. More significant features is that $|\nabla \rho|$ near the cloud boundary in the shocked gas are the highest. This is because the local droplet evaporation considerably alters the local gas density and temperature. Similar distributions can also be seen in Figs. 16(b) 一 16(d). The pressure gradient magnitude is high along the lead shock and also the transverse shocks extending from the triple points. However, no obvious misalignment of pressure and density gradients can be found at the cloud boundary, particularly at the southern/northern poles. This is different from the RMI problems in shock-bubble interactions, in which the baroclinity is highest at these two poles [35,50]. At 108 $\mu s$, $|\nabla \rho|$ near the southern/northern poles becomes high, but no high pressure gradient distributions exist there. Therefore, although the large-scale vortices are developing from the upper and lower boundaries as shown in Fig. 14, nonetheless, clear misalignment of pressure and density gradients does not occur. This can also be confirmed from our results of baroclinity at these instants, as shown in Fig. S6 of the supplementary document. At later instants (e.g., 116 and 154 $\mu s$), as the cloud evolves, the density gradient are still obvious at the cloud boundaries. The pressure fields becomes more complex, characterized by many shocklets (e.g., from transverse shocks) interacting with the cloud. This induces the formation of small-scale structures along the two-phase fluid interface, as shown in Fig. 14(1). This phenomenon has not been reported in the shock-cloud interactions studies [20,51] due to the relatively clear pressure field in a shocked flow.

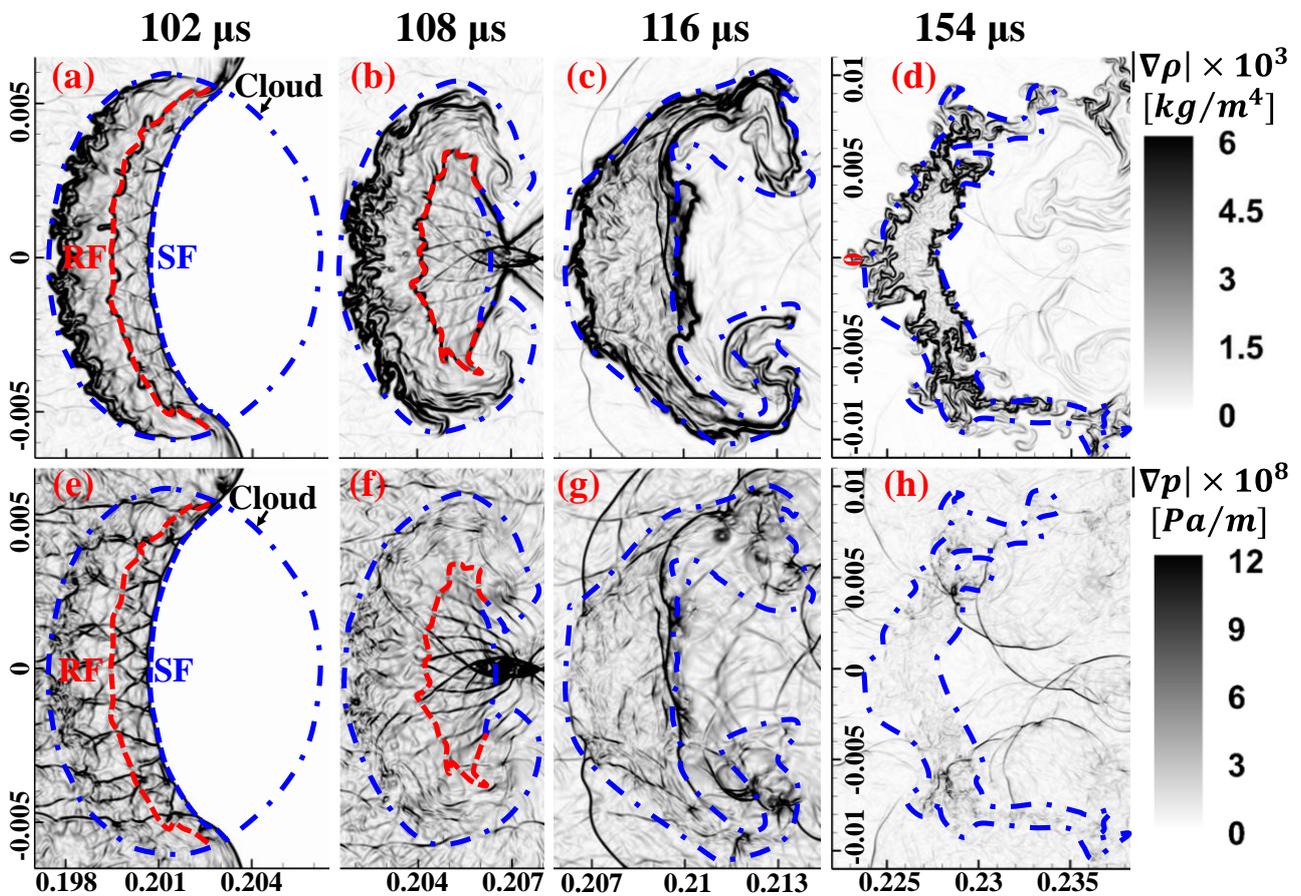

Figure 16 Time sequence of density (first row) and pressure (second row) gradient magnitudes. SF/RF: shock/reaction front. Dash-dotted line: cloud boundary. Axis label unit: m.

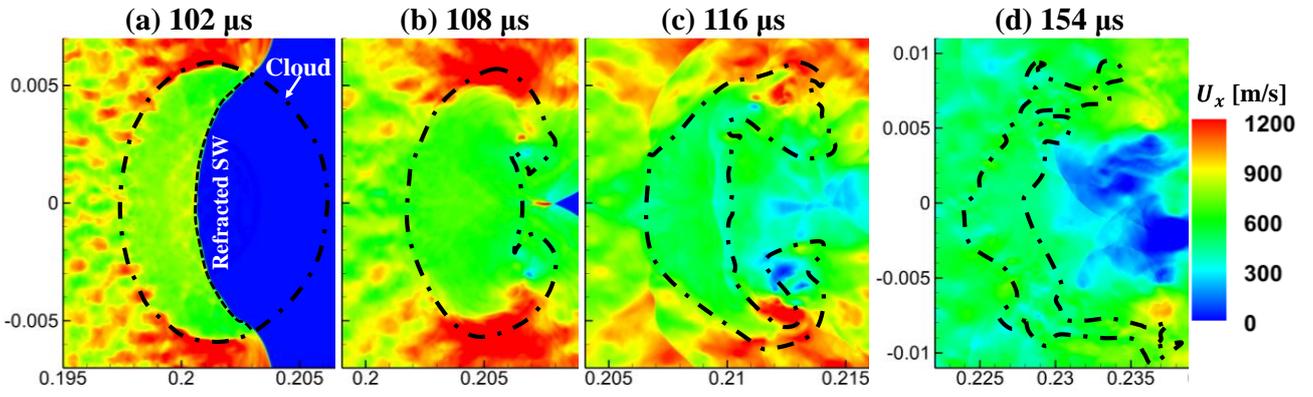

Figure 17 Distributions of *x*-component gas velocities. Dash-dotted line: cloud boundary. Axis label unit: m.

To further explore the mechanism of large-scale vortex roll-up from the upper and lower boundaries, we examine the distributions of *x*-component gas velocity in Fig. 17. It is seen that large gas velocity difference (up to 600 m/s) between the cloud and surrounding gas is observed, e.g., in Fig. 17(a). Due to existence of the static cloud, momentum absorption (for accelerating the droplets) between two phase causes lower gas speed in the cloud than the surrounding one. This difference results in considerable shear along the cloud boundary, particularly around the southern/northern poles, which leads to formation of the vortex roll-up observed in Fig. 15. However, at a later instant, like 154 $\mu s$, since the droplets kinematically equilibrate with the gas, the velocity difference along the fluid interface almost disappears, as shown in Fig. 17(d).

Therefore, although the vortex formation and development in Fig. 14 are superficially similar to those in the previous RMI problems (see detailed reviews in Refs.[50,53]), nonetheless, their underpinning mechanisms are different. Although Vorobieff et al. make some conjectures about this, they do not articulate in their work [52]. In the traditional RMI problems (like shock-bubble interaction), the density difference across the interface arises from the different gaseous materials and exists before the fluid is shocked. However, for the two-phase problems, gas phase properties are continuous across the interface before the shock or detonation wave arrives. The interfacial discontinuity continuously becomes strong when the interphase exchanges proceed in the cloud. Based on the above analysis, velocity difference (hence shear) dominates the initial development of vortices, which essentially corresponds to Kelvin–Helmholtz instability (KHI). Instead, RMI is not shown to play a significant role in the macroscopic evolution of the cloud, although the shocklets may interact with the cloud boundary where density gradient dominates. Therefore, in this work, the effective Atwood number, Eq. (8), more reflects the intensity of the two-phase interactions, instead of the density gradient across the fluid interface, as in traditional RMI problems. We also run a numerical experiment based on the case in Figs. 14-17, in which only momentum exchange is retained and the energy/mass exchanges are turned off. The results (see Fig. S7 of the supplementary document) shows that the vortex formation can still be clearly observed even if only the momentum exchange exists. This again confirms the role of the KHI in development of two-phase fluid interface.

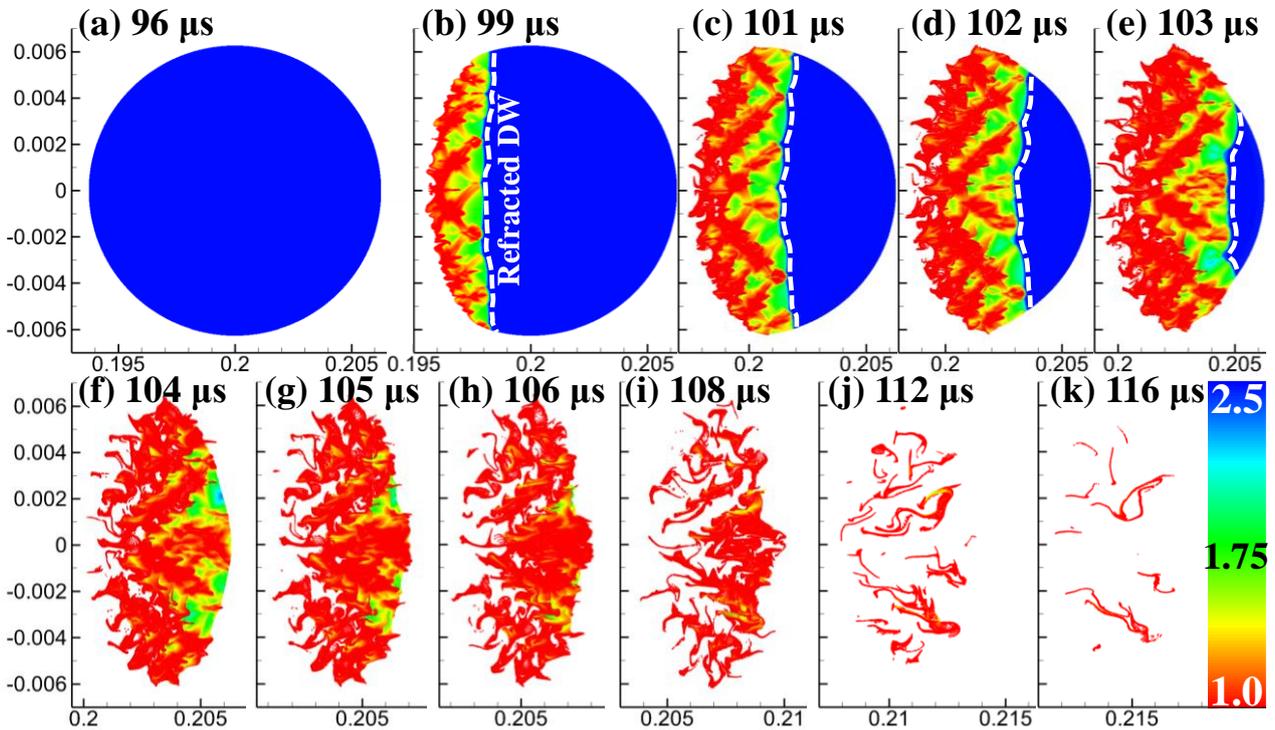

Figure 18 Evolutions of Lagrangian droplets colored by diameter (in $\mu m$). Dashed curve: refracted wave. $A_e = 0.11$, $c = 0.105$ kg/m$^3$, $d_d^0 = 2.5$ $\mu m$, and $R = 0.25W$. Axis label unit: m.

**4.4.2 Effective Atwood number effects**

The influences of effective Atwood number $A_e$ on the evolutions of cloud morphology are evaluated in Figs. 18—20. Figure 18 shows the cloud evolutions with a lower effective Atwood number, $A_e = 0.11$. The corresponding conditions are $c = 0.105$ kg/m³, $d_d^0 = 2.5$ $\mu m$, and $R = 0.25W$. Different from Fig. 14, small fingers-like structures develop from the cloud upstream boundary, and no droplet jets and tails can be observed. Due to the moving triple points, its high temperature and pressure leads to fragmentation of local droplets (smaller droplets along their tracks), leading to the cellularization of the shocked cloud. Such cloud behaviors are not seen from the preceding studies about multiphase instability [20,51,52], where only planar shocks are considered. Meanwhile, the droplet diameters behind the shock are reduced immediately behind the refracted shock, compared to those in Fig. 14. This is because with the same intensity of incident shock wave more droplets can be disintegrated when the water concentration is lower [5]. After the refracted shock wave leaves the cloud at 104 $\mu s$, almost all droplets are below 1.0 $\mu m$. Consequently, accelerated evaporation leads to quickly shrinking cloud. At 116 $\mu s$ the droplets are almost fully vaporized. In this case, due to relatively light loading of the droplets, limited influences on the gas field can be seen. For instance, no velocity difference can be observed between the cloud and surrounding gas from Fig. 21(a). Therefore, large-scale vortex does not appear from the south and north poles of the cloud, as shown in Fig. 21(d). This corresponds to the morphology of "cloud without tails", see Fig. 13.

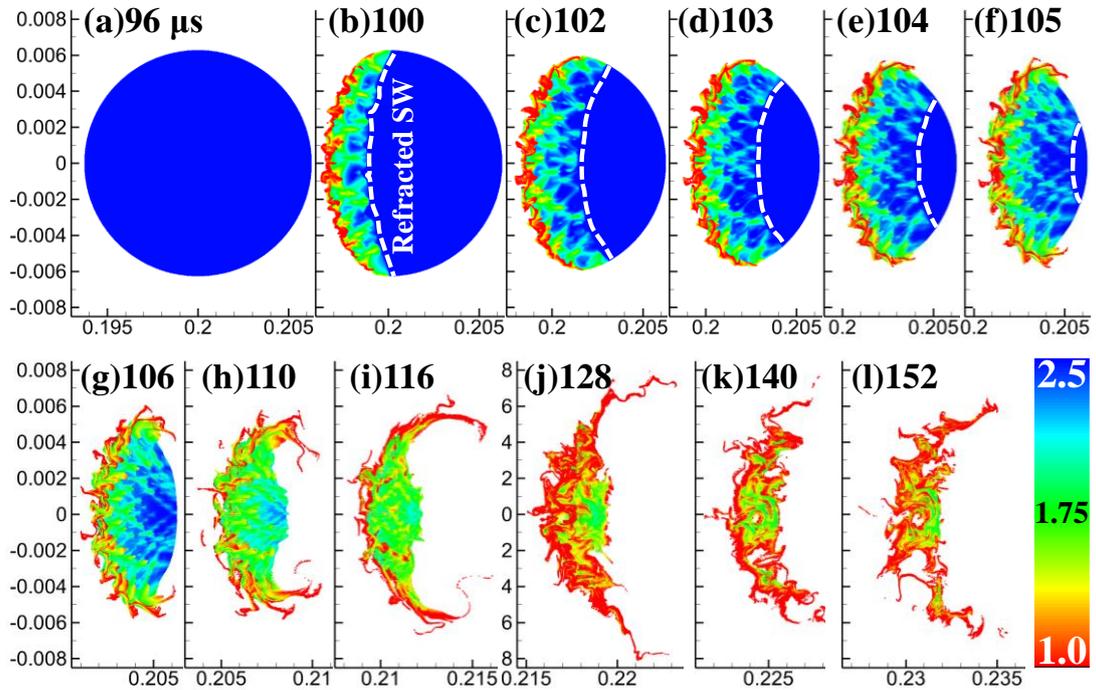

Figure 19 Evolutions of Lagrangian droplets colored by diameter (in $\mu m$). $c = 0.42$ kg/m³, $d_d^0 = 2.5$ $\mu m$, and $R = 0.25W$. Dashed curve: refracted shock wave. Axis label unit: m. $A_e = 0.33$.

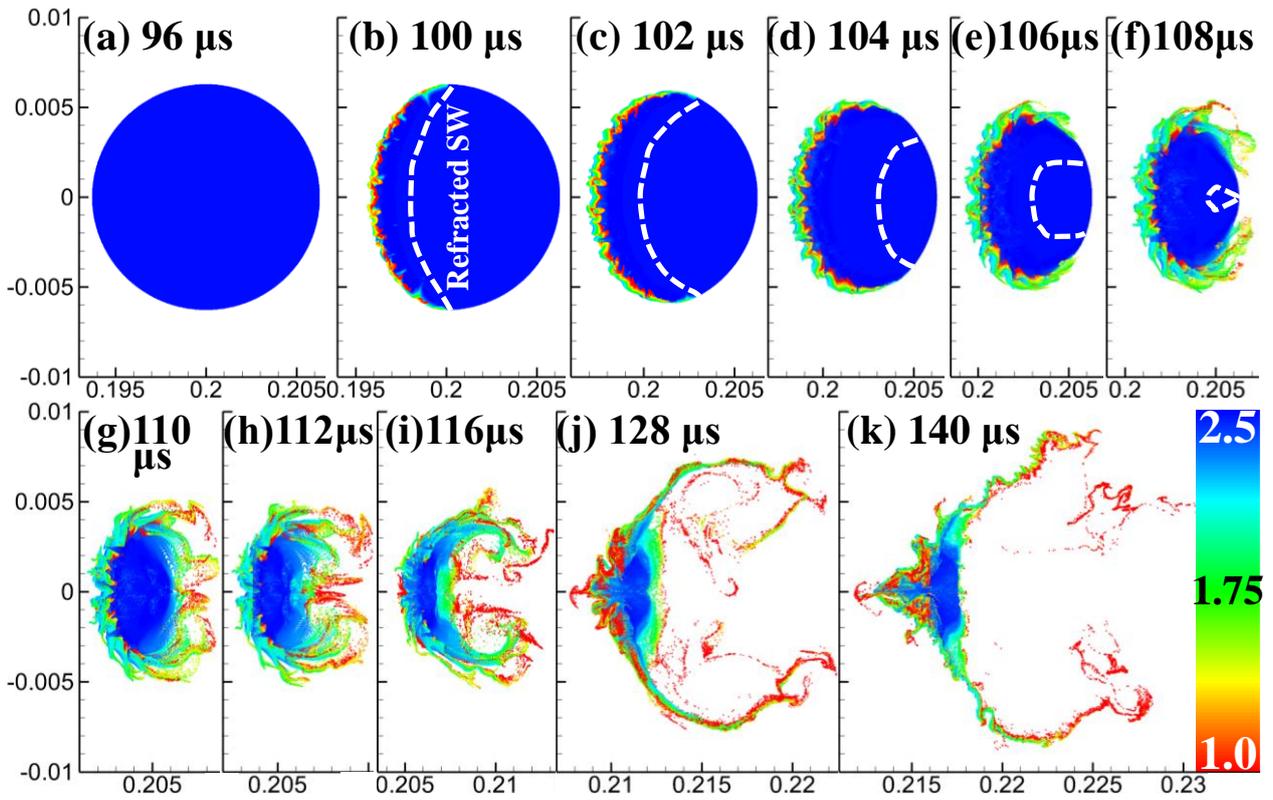

Figure 20 Evolutions of Lagrangian droplets colored by diameter (in $\mu m$). $c = 1.68$ kg/m$^3$, $d_d^0 = 2.5$ $\mu m$, and $R = 0.25W$. Dashed curve: refracted shock wave. Axis label unit: m. $A_e = 0.67$.

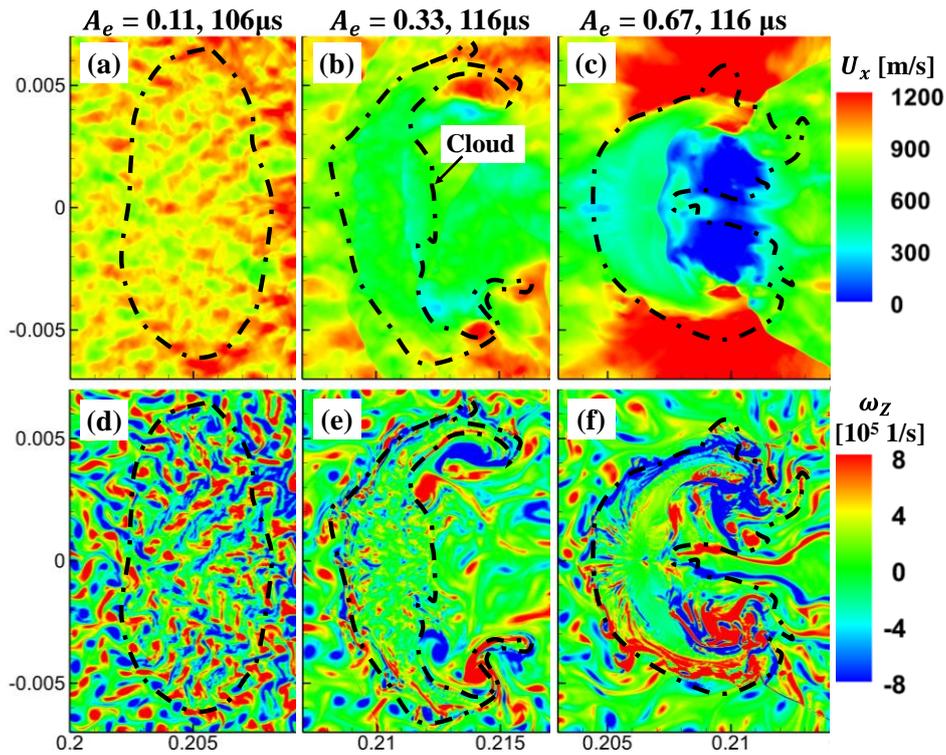

Figure 21 $z$-component vorticity and $x$-component velocity with different effective Atwood numbers. Dashed line: cloud. Axis label unit: m.

The evolutions of the cloud with an intermediate effective Atwood number, $A_e = 0.33$ ($c = 0.42$ kg/m$^3$), is demonstrated in Fig. 19. In the early instants, they have similar behaviors to those of $A_e = 0.5$ in Fig. 14. Nonetheless, some differences are observed after 116 $\mu s$: the cloud become thin, and the two tails from southern/northern poles are shorter. The z-component vorticity distribution in Fig. 21(e) confirms the occurrence of weaker vortex roll-up at 116 $\mu s$ due to velocity difference at the in Fig. 21(b). At later stage (e.g., 140 and 152 $\mu s$), the tails becomes shorter due to fast evaporation of the droplets. When $A_e = 0.67$, new cloud features can be found from Fig. 20. Specifically, at 110 — 116 $\mu s$ one can see the formation of relatively bigger counter-rotating vortex pair, which is also observed from Fig. 21(f). This is directly associated with the larger speed difference in Fig. 21(c).

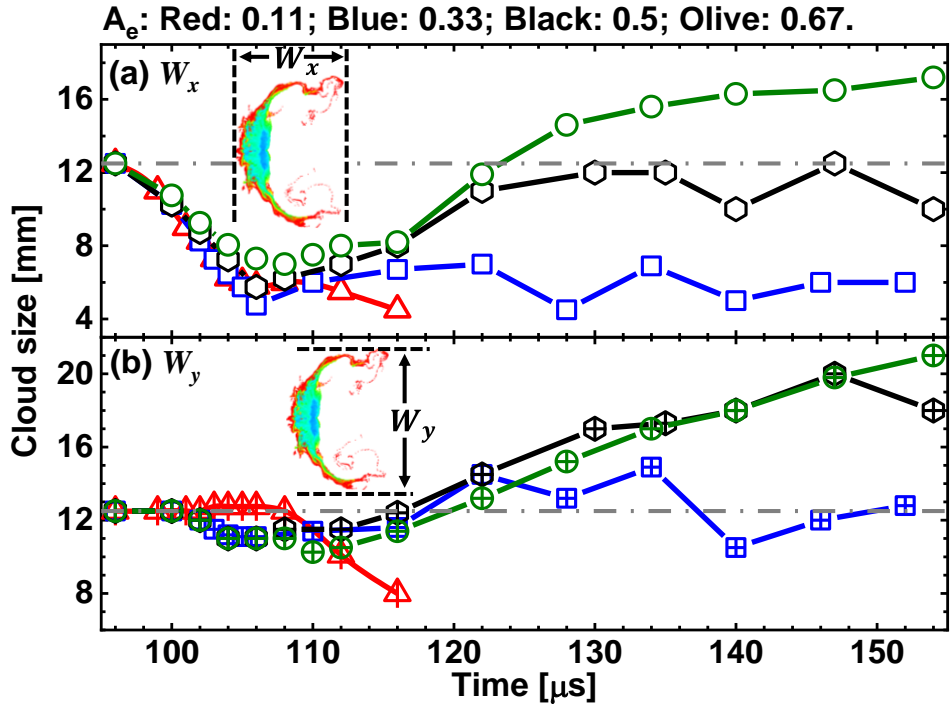

Figure 22 Time history of the cloud size in (a) x-direction $W_x$ and (b) y-direction $W_y$ with various $d_d^0 = 2.5$ $\mu m$ and $R = 0.25W$. Dashed line: initial cloud diameter.

The influences of effective Atwood number on the cloud evolutions are quantified through the change of mixing width in Fig. 22. Following Vorobieff et al. [52], the mixing extent is defined as the distance between the edges of the cloud along the x- / y-direction respectively, i.e., $W_x$ and $W_y$ in

Fig. 22. The edges are identified when the droplet volume fraction is greater than $1\times10^{-5}$. The initial cloud diameter is 12.5 mm. In Fig. 22(a), when $A_e = 0.11$, $W_x$ is monotonically reduced to approximately 4 mm until complete gasification at 116 $\mu s$. This is due to the movement of the upstream boundary after shock impacting, as seen from Fig. 18. However, when $A_e = 0.33$ and 0.5, after initial reduction similar to the $A_e = 0.11$ curve, $W_x$ increases at about 106 $\mu s$ and then levels off respectively around 6 and 10 mm, due to the growing cloud tails. Different from these two cases, continuously increased $W_x$ is observed with the greatest $A_e$ (0.67) after 110 $\mu s$. It is even larger than the initial cloud diameter. This increase is mainly caused by longer tails from larger velocity difference of the two fluids (see Fig. 21f, hence stronger vorticity).

In Fig. 22(b), when $A_e = 0.11$, the *y*-component cloud size $W_y$ is almost not changed before 106 $\mu s$, and thereafter quickly drops below 8 mm due to fast evaporation of the disintegrated droplets near the southern/northern poles. Nevertheless, when $A_e$ is beyond 0.11, $W_y$ slightly decreases and then becomes greater after 110 $\mu s$ due to the developing tails. Moreover, $W_y$ with $A_e = 0.33$ exhibits some fluctuations around 12.5 mm, but in the two larger $A_e$ cases it continuously increases. Overall, the results in Fig. 22 indicate that higher droplet concentration (higher $A_e$) would lead to wider droplet dispersion range due to formation of larger vortices. From practical measures for $H_2$ explosion inhibition, the spreading evaporating droplets can cool and dilute a larger extent of detonated area and therefore more effectively mitigate the possible secondary hazards behind the DW.

## 5. Conclusions

Interactions between a propagating detonation wave and circular water spray cloud in hydrogen/air mixture are simulated by Eulerian – Lagrangian approach and detailed chemical mechanism. Parametric studies are performed through considering a range of droplet diameter, concentration and cloud size.

It is seen that droplet size, concentration and cloud radius have significant effects on peak pressure trajectories of the detonation wave. Three detonation propagation modes are observed: perturbed propagation, leeward re-detonation, and extinction. Leeward re-detonation is analyzed with unsteady evolutions of gas phase and liquid phase quantities. The refracted detonation wave inside the cloud is decoupled and propagates more slowly than the outer detonation. Here the converging effects of the water cloud refract and diffract the incident DW. The re-initiation is caused by the shock focusing from upper and lower diffracted detonations. The results also show that breakup of water droplets proceeds when the detonation wave crosses the cloud.

Furthermore, detonation extinction is observed when we vary the water cloud size. It is featured by quickly fading peak pressure trajectories when the detonation passes the cloud, and no local autoignition occurs in the shock focusing area. Evolutions of thermochemical structures along the domain centerline between the reaction front and shock front in an extinction process are also studied. The transfer rates of mass, energy and momentum of detonation success and failure are analyzed. In addition, a series of cloud-gas interaction simulations considering various droplet concentrations and cloud radii are performed. It is shown that the critical cloud size to quench a detonation decreases when the droplet concentration is increased. However, when the droplet concentration is beyond 0.84 kg/m$^3$, this critical cloud size is almost not affected by it.

Two-phase interfacial instability is analyzed with various effective Atwood number, which results in two typical cloud evolutions: shrinking clouds without tails or with tails. Mechanism of droplet cloud evolution is demonstrated with the time history of Lagrangian water droplet distributions, vorticity, density / pressure gradient magnitudes and gas velocity. Our results confirm that velocity difference (shear) dominates the initial development of vortices, corresponding to Kelvin–Helmholtz instability. Instead, Richtmyer–Meshkov instability makes small contributions to macroscopic evolution of the cloud. Moreover, the influences of effective Atwood number on the evolutions of cloud morphology, vorticity, velocity, and cloud size are evaluated. Analysis shows that higher droplet concentration would lead to wider droplet dispersion range due to formation of larger vortices.

**Acknowledgements**

This work used the computational resources of ASPIRE 1 Cluster in The National Supercomputing Centre, Singapore (https://www.nscc.sg) and The RIKEN R-CCS Supercomputer Fugaku in Japan (hp210196). YX is supported by the NUS Research Scholarship. This research is supported by the National Research Foundation, Prime Minister's Office, Singapore under its Campus for Research Excellence and Technological Enterprise (CREATE) program (A-0009468-00-00).

**References**

[1] G. Grant, J. Brenton, D. Drysdale, Fire suppression by water sprays, Prog. Energy Combust. Sci. 26 (2000) 79–130.

[2] Y. Xu, M. Zhao, H. Zhang, Extinction of incident hydrogen/air detonation in fine water sprays, Phys. Fluids. 33 (2021) 116109.

[3] G. Jourdan, L. Biamino, C. Mariani, C. Blanchot, E. Daniel, J. Massoni, L. Houas, R. Tosello, D. Praguine, Attenuation of a shock wave passing through a cloud of water droplets, Shock Waves. 20 (2010) 285–296.

[4] A. Chauvin, G. Jourdan, E. Daniel, L. Houas, R. Tosello, Experimental investigation of the propagation of a planar shock wave through a two-phase gas-liquid medium, Phys. Fluids. 23 (2011) 1–13.

[5] K.C. Adiga, H.D. Willauer, R. Ananth, F.W. Williams, Implications of droplet breakup and formation of ultra fine mist in blast mitigation, Fire Saf. J. 44 (2009) 363–369.

[6] R. Ananth, H.D. Willauer, J.P. Farley, F.W. Williams, Effects of Fine Water Mist on a Confined Blast, Fire Technol. 48 (2012) 641–675.

[7] D.A. Schwer, K. Kailasanath, Numerical simulations of the mitigation of unconfined explosions using water-mist, Proc. Combust. Inst. 31 II (2007) 2361–2369.

[8] K. Shibue, Y. Sugiyama, A. Matsuo, Numerical study of the effect on blast-wave mitigation of the quasi-steady drag force from a layer of water droplets sprayed into a confined geometry, Process Saf. Environ. Prot. 160 (2022) 491–501.

[9] G.O. Thomas, M.J. Edwards, D.H. Edwards, Studies of detonation quenching by water sprays, Combust. Sci. Technol. 71 (1990) 233–245.

[10] U. Niedzielska, L.J. Kapusta, B. Savard, A. Teodorczyk, Influence of water droplets on propagating detonations, J. Loss Prev. Process Ind. 50 (2017) 229–236.

[11] G. Jarsalé, F. Virot, A. Chinnayya, Ethylene–air detonation in water spray, Shock Waves. 26 (2016) 561–572.

[12] H. Watanabe, A. Matsuo, A. Chinnayya, K. Matsuoka, A. Kawasaki, J. Kasahara, Numerical analysis of the mean structure of gaseous detonation with dilute water spray, J. Fluid Mech. 887 (2020) A4-1–40.

[13] H. Watanabe, A. Matsuo, A. Chinnayya, K. Matsuoka, A. Kawasaki, J. Kasahara, Numerical analysis on behavior of dilute water droplets in detonation, Proc. Combust. Inst. 000 (2020) 1–8.

[14] J. Shi, Y. Xu, W. Ren, H. Zhang, Critical condition and transient evolution of methane detonation extinction by fine water droplet curtains, Fuel. 315 (2022) 123133.

[15] Y. Xu, H. Zhang, Pulsating propagation and extinction of hydrogen detonations in ultrafine water sprays, Combust. Flame. 241 (2022) 112086.

[16] Z.Y. Zhou, S.B. Kuang, K.W. Chu, A.B. Yu, Discrete particle simulation of particle-fluid flow: Model formulations and their applicability, J. Fluid Mech. 661 (2010) 482–510.

[17] C.T. Crowe, J.D. Schwarzkopf, M. Sommerfeld, Y. Tsuji, Multiphase flows with droplets and particles, CRC Press, New York, U.S., 1998.

[18] M. Pilch, C.A. Erdman, Use of breakup time data and velocity history data to predict the maximum size of stable fragments for acceleration-induced breakup of a liquid drop, Int. J. Multiph. Flow. 13 (1987) 741–757.

[19] L. Craig, A. Moharreri, A. Schanot, D.C. Rogers, B. Anderson, S. Dhaniyala, Characterizations of cloud droplet shatter artifacts in two airborne aerosol inlets, Aerosol Sci. Technol. 47 (2013) 662–671.

[20] V. Duke-Walker, W.C. Maxon, S.R. Almuhna, J.A. McFarland, Evaporation and breakup effects in the shock-driven multiphase instability, J. Fluid Mech. 908 (2021) 1–27.

[21] A.B. Liu, D. Mather, R.D. Reitz, Modeling the effects of drop drag and breakup on fuel sprays, SAE Tech. Pap. 1 (1993).

[22] W.E. Ranz, W.R. Marshall, Evaporation from Drops, Part I., Chem. Eng. Prog. 48 (1952) 141–146.

[23] C.T. Crowe, M.P. Sharma, D.E. Stock, The particle-source-in cell (PSI-CELL) model for gas-droplet flows, J. Fluids Eng. 99 (1977) 325–332.

[24] Z. Huang, M. Zhao, Y. Xu, G. Li, H. Zhang, Eulerian-Lagrangian modelling of detonative combustion in two-phase gas-droplet mixtures with OpenFOAM: Validations and verifications, Fuel. 286 (2021) 119402.

[25] M. Zhao, Z. Ren, H. Zhang, Pulsating detonative combustion in n-heptane/air mixtures under off-stoichiometric conditions, Combust. Flame. 226 (2021) 285–301.

[26] M. Zhao, M.J. Cleary, H. Zhang, Combustion mode and wave multiplicity in rotating detonative combustion with separate reactant injection, Combust. Flame. 225 (2021) 291–304.

[27] M. Zhao, H. Zhang, Large eddy simulation of non-reacting flow and mixing fields in a rotating detonation engine, Fuel. 280 (2020) 118534.

[28] Z. Huang, H. Zhang, On the interactions between a propagating shock wave and evaporating water droplets, Phys. Fluids. 32 (2020) 123315.

[29] A. Kurganov, S. Noelle, G. Petrova, Semidiscrete central-upwind schemes for hyperbolic conservation laws and Hamilton-Jacobi equations, 2002.

[30] M.P. Burke, M. Chaos, Y. Ju, F.L. Dryer, S.J. Klippenstein, Comprehensive H2/O2 kinetic model for high-pressure combustion, Int. J. Chem. Kinet. 44 (2012) 444–474.

[31] S. Yungster, K. Radhakrishnan, Pulsating one-dimensional detonations in hydrogen-air mixtures, Combust. Theory Model. 8 (2004) 745–770.

[32] M. Sontheimer, A. Kronenburg, O.T. Stein, Grid dependence of evaporation rates in Euler–Lagrange simulations of dilute sprays, Combust. Flame. 232 (2021) 111515.

[33] Y. Ling, J.L. Wagner, S.J. Beresh, S.P. Kearney, S. Balachandar, Interaction of a planar shock wave with a dense particle curtain: Modeling and experiments, Phys. Fluids. 24 (2012).

[34] M. Sommerfeld, The unsteadiness of shock waves propagating through gas-particle mixtures, Exp. Fluids. 3 (1985) 197–206.

[35] D. Ranjan, J. Oakley, R. Bonazza, Shock-bubble interactions, Annu. Rev. Fluid Mech. 43 (2011) 117–140.

[36] P.E. Dimotakis, R. Samtaney, Planar shock cylindrical focusing by a perfect-gas lens, Phys. Fluids. 18 (2006) 9–12.

[37] F. Bartlmä, K. Schröder, The diffraction of a plane detonation wave at a convex corner, Combust. Flame. 66 (1986) 237–248.

[38] F. Pintgen, J.E. Shepherd, Detonation diffraction in gases, Combust. Flame. 156 (2009) 665–677.

[39] Y. Nagura, J. Kasahara, Y. Sugiyama, A. Matsuo, Comprehensive visualization of detonation-diffraction structures and sizes in unstable and stable mixtures, Proc. Combust. Inst. 34 (2013) 1949–1956.

[40] Y. Nagura, J. Kasahara, A. Matsuo, Multi-frame visualization for detonation wave diffraction, Shock Waves. 26 (2016) 645–656.

[41] R. Samtaney, N.J. Zabusky, Circulation Deposition on Shock-Accelerated Planar and Curved

Density-Stratified Interfaces: Models and Scaling Laws, J. Fluid Mech. 269 (1994) 45–78.

[42]  J. Ou, J. Ding, X. Luo, Z. Zhai, Effects of Atwood number on shock focusing in shock–cylinder interaction, Exp. Fluids. 59 (2018) 1–11.

[43]  N. Haehn, D. Ranjan, C. Weber, J. Oakley, D. Rothamer, R. Bonazza, Reacting shock bubble interaction, Combust. Flame. 159 (2012) 1339–1350.

[44]  T.F. Lu, C.S. Yoo, J.H. Chen, C.K. Law, Three-dimensional direct numericaal simulation of a turbulent lifted hydrogen jet flame in heated coflow: A chemical explosive mode analysis, J. Fluid Mech. 652 (2010) 45–64.

[45]  D.A. Goussis, H.G. Im, H.N. Najm, S. Paolucci, M. Valorani, The origin of CEMA and its relation to CSP, Combust. Flame. 227 (2021) 396–401.

[46]  W. Wu, Y. Piao, Q. Xie, Z. Ren, Flame diagnostics with a conservative representation of chemical explosive mode analysis, AIAA J. 57 (2019) 1355–1363.

[47]  F. Diegelmann, S. Hickel, N.A. Adams, Three-dimensional reacting shock–bubble interaction, Combust. Flame. 181 (2017) 300–314.

[48]  F. Diegelmann, V. Tritschler, S. Hickel, N. Adams, On the pressure dependence of ignition and mixing in two-dimensional reactive shock-bubble interaction, Combust. Flame. 163 (2016) 414–426.

[49]  F. Diegelmann, S. Hickel, N.A. Adams, Shock Mach number influence on reaction wave types and mixing in reactive shock–bubble interaction, Combust. Flame. 174 (2016) 85–99.

[50]  M. Brouillette, The richtmyer-meshkov instability, Annu. Rev. Fluid Mech. 34 (2002) 445–468.

[51]  J.B. Middlebrooks, C.G. Avgoustopoulos, W.J. Black, R.C. Allen, J.A. McFarland, Droplet and multiphase effects in a shock-driven hydrodynamic instability with reshock, Exp. Fluids. 59 (2018) 1–16.

[52]  P. Vorobieff, M. Anderson, J. Conroy, R. White, C.R. Truman, S. Kumar, Vortex formation in a shock-accelerated gas induced by particle seeding, Phys. Rev. Lett. 106 (2011) 1–4.

[53]  N.J. Zabusky, Vortex paradigm for accelerated inhomogeneous flows: Visiometrics for the Rayleigh-Taylor and Richtmyer-Meshkov Environments, Annu. Rev. Fluid Mech. 31 (1999) 495–536.